\documentstyle[aps,multicol,tighten,eqsecnum,array,psfig]{revtex}
\begin{document}
\draft
\title{Atom Lasers, Coherent States, and Coherence: \\ I. Physically
Realizable Ensembles of Pure States}
\author{H.M. Wiseman$^{1,2,3,4}$\footnote{Electronic address:
h.wiseman@gu.edu.au}
and John A. Vaccaro$^{2,4,1}$}
\address{$^{1}$School of Science, Griffith University, Brisbane 4111
Australia.}
\address{$^{2}$Division of Physics and Astronomy, University of
Hertfordshire, Hatfield AL10 9AB, UK.}
\address{$^{3}$Department of Physics, University of Queensland, Queensland
4072 Australia.}
\address{$^{4}$Physics Department, The Open University, Milton Keynes
MK7 6AA,  United Kingdom}

\maketitle

\begin{center}(\today)\end{center}

\begin{abstract}
A laser, be it an optical laser or an atom laser, is an open
quantum system that produces a coherent beam of bosons (photons or
atoms respectively). Far above threshold, the stationary state
$\rho_{\rm ss}$ of the laser mode is a mixture of coherent field
states with random phase, or, equivalently, a Poissonian mixture
of number states. This paper answers the question: can
descriptions such as these, of $\rho_{\rm ss}$ as a stationary
ensemble of pure states, be physically realized? Here physical
realization is as defined previously by us [H.M. Wiseman and J.A.
Vaccaro, Phys. Lett. A {\bf 250}, 241 (1998)]: an ensemble of pure
states for a particular system can be physically realized if,
without changing the dynamics of the system, an experimenter can
(in principle) know at any time that the system is in one of the
pure-state members of the ensemble. Such knowledge can be obtained
by monitoring the baths to which the system is coupled, provided
that coupling is describable by a Markovian master equation. Using
a family of master equations for the (atom) laser, we solve for
the physically realizable (PR) ensembles. We find that for any
finite self-energy $\chi$ of the bosons in the laser mode, the
coherent state ensemble is not PR; the closest one can come to it
is an ensemble of squeezed states. This is particularly relevant
for atom lasers, where the self-energy arising  from elastic
collisions is expected to be large. By contrast, the number state
ensemble is always PR. As the self-energy $\chi$ increases, the
states in the PR ensemble closest to the coherent state ensemble
become increasingly squeezed. Nevertheless, there are values of
$\chi$ for which states with well-defined coherent amplitudes are
PR, even though the atom laser is not coherent (in the sense of
having a Bose-degenerate output). We discuss the physical
significance of this anomaly in terms of conditional coherence
(and hence conditional Bose degeneracy).
\end{abstract}


\newcommand{\beq}{\begin{equation}}
\newcommand{\eeq}{\end{equation}}
\newcommand{\bqa}{\begin{eqnarray}}
\newcommand{\eqa}{\end{eqnarray}}
\newcommand{\nn}{\nonumber}
\newcommand{\nl}{\nn \\ &&}
\newcommand{\dg}{^\dagger}
\newcommand{\erf}[1]{Eq.~(\ref{#1})}
\newcommand{\smallfrac}[2]{\mbox{$\frac{#1}{#2}$}}
\newcommand{\bra}[1]{\left\langle{#1}\right|}
\newcommand{\ket}[1]{\left|{#1}\right\rangle}
\newcommand{\ip}[2]{\left\langle{#1}\right|\left.{#2}\right\rangle}
\newcommand{\sch}{Schr\"odinger }
\newcommand{\schs}{Schr\"odinger's }
\newcommand{\hei}{Heisenberg }
\newcommand{\heis}{Heisenberg's }
\newcommand{\half}{\smallfrac{1}{2}}
\newcommand{\bl}{{\bigl(}}
\newcommand{\br}{{\bigr)}}
\newcommand{\ito}{It\^o }
\newcommand{\sq}[1]{\left[ {#1} \right]}
\newcommand{\cu}[1]{\left\{ {#1} \right\}}
\newcommand{\ro}[1]{\left( {#1} \right)}
\newcommand{\an}[1]{\left\langle{#1}\right\rangle}
\newcommand{\implies}{\Longrightarrow}
\newcommand{\tr}[1]{{\rm Tr}\sq{ {#1} }}
\newcommand{\st}[1]{\left| {#1} \right|}
\newcommand{\singlecol}{\end{multicols}
     \vspace{-0.5cm}\noindent\rule{0.5\textwidth}{0.4pt}\rule{0.4pt}
     {\baselineskip}\widetext }
\newcommand{\doublecol}{\noindent\hspace{0.5\textwidth}
     \rule{0.4pt}{\baselineskip}\rule[\baselineskip]
     {0.5\textwidth}{0.4pt}\vspace{-0.5cm}\begin{multicols}{2}\noindent}

\begin{multicols}{2}
\narrowtext

\section{Introduction}

In elementary presentations of quantum optics it is more or less
an axiom that a laser field is represented by a coherent state
$\ket{\alpha}$. Recently, it has been argued that this
representation is a fiction, albeit a convenient one \cite{Mol96}.
The essential argument is that no commonly employed process at
optical frequencies produces an electric field having a non-zero
average amplitude. While this point of view is certainly
defensible \cite{Gea98Mol98}, it perhaps obscures the fact that
there is something special about laser light.

In Ref.~\cite{Wis97}, one of us argued that what is special about
laser light is that it is well approximated by a noiseless
classical electromagnetic wave. Four quantitative criteria were
given, none of which require a mean field, so there is no dispute
with Ref.~\cite{Mol96}. The least familiar, and so most
important, of these criteria is that the output flux of the laser
(bosons per unit time) must be much greater than its spectral
linewidth. Put another way, the coherence time of a true laser
must be much greater than the mean temporal separation of photons
in the output beam. This is typically satisfied by many orders of
magnitude in optical lasers, but is not satisfied by ordinary
thermal sources.

This concept of quantum coherence is quite distinct from the
elementary idea that a laser is in a coherent state. Indeed, it is
compatible with theoretical models for typical laser processes
\cite{SarScuLam74,Lou73}, which imply that the state of the cavity
mode for a laser far above threshold is a mixture of coherent
states of all phases. That is to say, the stationary state matrix
of the laser mode can be written
\beq
\rho_{\rm ss} = \int \frac{d\phi}{2\pi}
\ket{|\alpha|e^{i\phi}}\bra{|\alpha|e^{i\phi}}, \label{mixcoh}
\eeq
where $|\alpha|^{2}=\mu$ is the mean number of photons in the laser.

It would be tempting to interpret \erf{mixcoh} to mean that the laser
really is in a coherent state $\ket{|\alpha|e^{i\phi}}$ of definite
phase $\phi$, but we don't know what that phase is. However, this
temptation must be resisted because the stationary state matrix can
also be written
\beq
\rho_{\rm ss} = \sum_{n=0}^{\infty}
e^{-\mu}\frac{\mu^{n}}{n!}\ket{n}\bra{n} ,\label{mixnum}
\eeq
which would seem to imply that the laser really is in a number
state $\ket{n}$, but we don't know which number it is.

The ``unknown coherent state'' description and the ``unknown
number state'' description are {\em mathematically equivalent}
representations of the stationary state matrix $\rho_{\rm ss}$.
However, in the physical context that $\rho_{\rm ss}$ is the
stationary state of an open quantum system in {\em dynamical
equilibrium}, the two representations are {\em not physically
equivalent}. This idea is at the heart of this paper and the
following paper \cite{WisVac01b}. In this paper we investigate
whether these, and other pure state ensembles are {\em physically
realizable}. We will show that under some circumstances, the
``unknown coherent state'' description is not physically
realizable, in contrast to the ``unknown number state''
description, which is. In the following paper we look at the
question of the how {\em robust} the ensembles are. We find that
even among physically realizable ensembles, a physical distinction
may be drawn based upon the {\em survival time}, the average time
that a member of the ensemble remains close to its original state
when left to evolve under the system dynamics. Both of these
concepts, the physical realizability of pure-state ensembles, and
the robustness of such ensembles, were introduced in an earlier
work by us \cite{WisVac98}.

Before proceeding further, it is necessary to clarify what we mean
by ``physically realizable'' (PR). A stationary pure state
ensemble of a given system is PR if it is possible, {\em without
altering the dynamics of the system}, to know that its state at
equilibrium is definitely one of the pure states in the ensemble.
Of course {\em which} pure state cannot be predicted beforehand.
It may seem contradictory to say that the system at equilibrium is
mixed, but that nevertheless we can know it to be in a pure state.
The resolution is that, by monitoring the system's environment,
the system state can, under suitable circumstances, be collapsed
over time into a pure state. Being simply an example of a quantum
measurement, this process, called an {\em unraveling}
\cite{Car93b}, will be stochastic. {\em On average}, the system
evolution is not changed and the ensemble of pure states produced
by the unraveling is guaranteed to be equivalent to the
equilibrium mixed state.

From this description it should be apparent that the question of
whether an ensemble is PR or not cannot be determined from the
stationary mixed state $\rho_{\rm ss}$. Rather, it depends upon
the dynamics (reversible and irreversible) that produced the
stationary state. Indeed, the unraveling to a pure state is
realized by monitoring the environment of the system, the same
environment that produces the irreversible dynamics of the system.
It would not be justifiable to introduce some new reservoir to
allow a new measurement to be made. Even if that did not change
the stationary state of the system (such as would be the case for
adding a QND-boson number measurement to a laser), it would change
the dynamics of the system, and hence one would be investigating a
{\em different system}.

The fact that different dynamics can lead to the same stationary
mixed state is easy to see for the case of a laser. Any process
that commutes with boson number will not alter the stationary
laser state $\rho_{\rm ss}$, since its eigenstates are the number
states, as shown by \erf{mixnum}. An example of an irreversible
process that commutes with boson number is phase diffusion. This
is relevant to all current lasers, which have some phase diffusion
in excess of the standard limit (although see Ref.~\cite{Wis99b}
for theoretical proposals for lasers that have phase diffusion
below the standard limit). There are also reversible processes
that commute with boson number, such as degenerate four-wave
mixing. While this dynamics is unimportant in most optical lasers,
it is expected to be very significant in {\em atom lasers}.

An atom laser is a device that produces an output beam of bosonic
atoms analogous to an optical laser's beam of photons
\cite{Wis97}. The idea for an atom laser was published
independently by a number of authors
\cite{WisCol95,SprPfaJanWil95,OlsCasDal95,Hol96}, shortly after
the first achievement of Bose-Einstein condensation (BEC) of a
dilute atomic gas \cite{And95,Bra95,Dav95}.  There have since been
some important experimental advances in the coherent release of
pulses \cite{Mew97,And98} and beams \cite{Hag99,Blo99} of atoms
from a condensate. Because the condensate is not replenished in
these experiments, the output coupling cannot continue
indefinitely, so these devices are only the first steps towards
achieving a CW atom laser.

Even though the atoms in the current BEC experiments are weakly
interacting in the sense of forming a gas rather than a liquid,
elastic collisions may dominate the dynamics of the condensate.
This self-interaction does not directly alter the number of atoms
in the condensate, and is analogous to four-wave mixing, (that is,
a $\chi^{(3)}$ nonlinearity), in optics. In this paper we show
that the presence of this nonlinearity has an enormous influence
on what ensembles of pure states are physically realizable. It
also determines the laser linewidth, and in this paper, we explore
the connection, between the presence of a PR coherent amplitude,
and the coherence of the laser output.

This paper is organized as follows. In Sec.~II we explain in
detail our concept of physically realizable pure-state ensembles
for open quantum systems. In Sec.~III we present our atom laser
model, including self-interactions and phase diffusion. In Sec.~IV
we apply the formalism of Sec.~II to the atom laser model and set
up the framework for calculating the PR ensembles. We calculate
the PR ensembles in Sec.~V and derive various scaling laws for the
ensembles as a function of the self interaction and phase
diffusion. We conclude in Sec.~VI with a summary and a discussion
of the interpretation and implications of our work.

\section{Physically Realizable Ensembles}

\subsection{The Master Equation}

Open quantum systems generally become entangled with their
environment, and this causes their state to become mixed. In many
cases, the system will reach an equilibrium mixed state in the
long time limit. A CW laser or atom laser is a system of this
sort, and we will restrict our consideration to such systems.  It
is common to refer to the environment of these systems as a
reservoir and, accordingly, we use both terms (environment and
reservoir) interchangeably here.

If the system is weakly coupled to the environmental reservoir,
and many modes of the reservoir are roughly equally affected by
the system, then one can make the Born and Markov approximations
in describing the effect of the environment on the system
\cite{Gar91}. Tracing over (that is, ignoring) the state of the
environment leads to a Markovian evolution equation for the state
matrix $\rho$ of the system, known as a {\em quantum master
equation}. The most general form of the quantum master equation
that is logically valid is the Lindblad form \cite{Lin76}
\beq
   \dot{\rho}= -i[H,\rho] + \sum_{k=1}^K {\cal D}[c_k]{\rho} \equiv
   {\cal L}\rho ,
   \label{genme}
\eeq
where for arbitrary operators $A$ and $B$,
\beq
   {\cal D}[A]B \equiv ABA\dg - \{A\dg A,B\}/2. \label{defcalD}
\eeq

If the master equation has a unique stationary state (as we will
assume it does), then that is defined by
\beq
{\cal L}\rho_{\rm ss} = 0.
\eeq
This assumption requires that ${\cal L}$ be time-independent. In
many quantum optical situations, one is only interested in the
dynamics in the interaction picture, in which the free evolution
at optical frequencies is removed from the state matrix. Indeed,
for quantum systems driven by a classical field, it may be
necessary to move into such an interaction picture in order to
obtain a time-independent Liouvillian superoperator ${\cal L}$.

The stationary state matrix $\rho_{\rm ss}$ can be expressed as an
ensemble of pure states as follows:
\beq
\rho_{\rm ss} = \sum_{n} \wp_{n} P_{n},
\eeq
where the $P_{n}$ are rank-one projection operators
\beq
P_{n} = \ket{\psi_{n}}\bra{\psi_{n}},
\eeq
and the $\wp_{n}$ are positive weights summing to unity. The
(possibly infinite) set of ordered pairs,
\beq
E = \{ (P_{n},\wp_{n}) :n=1,2,\ldots \},
\eeq
we will call an ensemble $E$ of pure states. Note that there is no
restriction that the projectors $P_{n}$ be mutually orthogonal.
This means that there are continuously infinitely many ensembles
$E$ that represent $\rho_{\rm ss}$. As noted in the introduction,
only some of these are physically realizable.

\subsection{Unravelings}

In the situation where a Markovian master equation can be derived,
it is possible (in principle) to continually measure the state of
the environment on a time scale large compared to the reservoir
correlation time but small compared to the response time of the
system. This effectively continuous measurement is what we will
call ``monitoring''. In such systems, monitoring the environment
does not disrupt the system--reservoir coupling and the system
will continue to evolve according to the master equation if one
ignores the results of the monitoring.

By contrast, if one does take note of the results of monitoring
the environment, then the system will no longer obey the master
equation (except on average). Because the system--reservoir
coupling causes the reservoir to become entangled with the system,
measuring the former's state yields information about the
latter's state. This will tend to undo the increase in the
mixedness of the system's state caused by the coupling.

If one is able to make perfect rank-one projective (i.e. von
Neumann \cite{Von32}) measurements of the reservoir state, with negligible time
delay from when it interacted with the system, then the system state
will usually be collapsed towards a pure state. However this is
not a process that itself can be described by projective
measurements on the system, because the system is not being
directly measured. Rather, the monitoring of the environment
leads to a gradual (on average) decrease in the system's entropy.

If the system is initially in a pure state then, under perfect
monitoring of its environment, it will remain in a pure state.
Then the effect of the monitoring is to cause the system to change
its pure state in a stochastic and (in general) nonlinear way.
Such evolution has been called a quantum trajectory \cite{Car93b},
and can be described by a nonlinear stochastic \sch equation
\cite{DalCasMol92,GarParZol92,WisMil93c}. The nonlinearity and
stochasticity are present because they are a fundamental part of
measurement in quantum mechanics.

Although a stochastic \sch equation is conceptually the simplest
way to define a quantum trajectory, in this work we will instead
use the stochastic master equation (SME) \cite{Bel88,BelSta92,%
Bar90,Bar93,WisMil93a}.

This has four general advantages. First, it can describe the
purification of an initially mixed state. Second, it can easily be
generalized to describe the situation where not all baths are
monitored perfectly, and the conditioned state never becomes pure
(as we will consider in Sec.~VI). Third, it is easier to see the
relation between the quantum trajectories and the master equation
that the system still obeys on average. Fourth, the form of the
SME is invariant under stochastic $U(1)$ transformations of the
state vector, which can radically alter the appearance (but not
the substance) of the stochastic \sch equation \cite{WisDio01}.

Assuming that the initial state of the system is pure, the quantum
trajectory for its projector $P$ will be described by the SME
\beq
  d{P}=  dt\left[ {\cal L} + {\cal U}(t) \right]P.
  \label{SSE1}
\eeq
Here ${\cal L}$ is the Liouvillian superoperator from the master
equation, and ${\cal U}$ is a stochastic superoperator which is,
in general, nonlinear in its operation on $P$. It also depends on
the operators $c_k$ as defined in \erf{genme}, and is constrained
by the following two equations, which must hold for arbitrary
rank-one projectors $P$
\bqa
\{P,({\cal L} + {\cal U})P\} + dt [{\cal U}P ][{\cal U}P ] &=&
({\cal L} + {\cal U})P,
 \label{prop1}\\
{\rm E}[{\cal U}P]&=& 0.    \label{prop2}
\eqa

The first of these properties ensures that $P+dP$ is also a
rank-one projector; that is, that the state remains pure. The
second ensures that
\beq d {\rm E}[P] = {\cal L}{\rm E}[P] dt,
\label{ensav} \eeq
where ${\rm E}$ denotes the ensemble-average
with respect to the stochasticity of ${\cal U}$. This
stochasticity is evidenced by the necessity of retaining the term
$dt [{\cal U}P][{\cal U}P ]$ in Eq.~(\ref{prop1}).

Because the ensemble average of the system still obeys the master
equation, the stochastic master equation (or equivalently the
stochastic \sch equation) is said to {\em unravel} the master
equation \cite{Car93b}. It is now well-known \cite{QSOSQO96} that
there are many (in fact continuously many) different unravelings
for a given master equation, corresponding to different ways of
monitoring the environment.

For simplicity we will call ${\cal U}$ an unraveling. Each
unraveling gives rise to an ensemble of pure states
\beq
   E^{\cal U} = \{ (P_{n}^{\cal U}, \wp_{n}^{\cal  U})
   :n=1,2,\ldots \},
   \label{ens}
\eeq
where $P_{n}^{\cal U}$ are the possible pure states of the system at
steady state, and $\wp_{n}^{\cal U}$ are their weights. For master
equations with   a unique stationary state $\rho_{\rm ss}$, the
SME (\ref{SSE1}) is ergodic over $E^{\cal U}$ and
$\wp_{n}^{\cal   U}$ is equal to the proportion of time the system
spends in state $P_{n}^{\cal U}$. The ensemble $E^{\cal U}$
represents $\rho_{\rm ss}$ in that
\beq
  \sum_{n} \wp_n^{\cal U} P_n^{\cal   U} = \rho_{\rm ss},
  \label{decomp}
\eeq
as guaranteed by Eq.~(\ref{ensav}).

\subsection{Continuous Markovian Unravelings}\label{seccmu}

To determine whether an ensemble $E$ is a PR ensemble $E^{\cal U}$
requires a search through the set, call it $J$, of all possible
unravelings ${\cal U}$. This set is extremely large. Although the
stochasticity in the superoperators ${\cal U}$ can always be
written in terms of quantum jumps, these jumps range in size from
being infinitesimal, to being so large that the system state after
the jump is always orthogonal to that before the jump
\cite{RigGis96}.

Another complication is that the unraveling need not be Markovian,
even though the master equation is. It might be thought that the
measurement must be Markovian since it must obtain full
information from the environment immediately after it has
interacted with the system  in order that the conditioned system
state remain pure. This rules out spectral detection, for example,
where the conditioned system state is not pure because it is
entangled with the state of the spectral filters \cite{WisToo99}.
However, the way in which the measurement obtains information from
the environment may not be independent of the past history of the
system. For example, the parameters defining the measurement may
depend on previous measurement results, resulting in an {\em
adaptive} measurement, as discussed in Ref.~\cite{Wis96a}.

From these considerations we see that a search over all possible
unravelings would not be practical. Thus it is useful, to consider
a smaller (but still continuously infinite) set $D$ containing
only unravelings that are continuous and Markovian (CM). A
continuous (but not differentiable) time evolution arises from
infinitely small (and infinitely frequent) jumps
\cite{Car93b,Wis96a}. In this case the probability distribution
for the pure states obeying the SME satisfies a Fokker-Planck
equation. On this basis it has been argued that these unravelings
are the natural ones to consider for quantum systems expected to
show quasi-classical behavior \cite{RigGis96}. The measurement
will be Markovian provided the measurement parameters $u_{jk}$
(defined below) are constants.

For the general master equation (\ref{genme}) the elements ${\cal
U}$ of $D$ can be written as \cite{WisVac98,WisDio01}
\beq
   {\cal U}(t) dt = \sum_{k=1}^K {\cal H}[dW^*_k(t) c_k].
\eeq
Here ${\cal H}[A]$ is a nonlinear superoperator defined, for
arbitrary operators $A$ and $B$, by
\beq
   {\cal H}[A]B \equiv AB + BA\dg - {\rm Tr}[AB + BA\dg]B,
   \label{defcalH}
\eeq
and the $dW_k(t)$ are the infinitesimal increments of a complex
multi-dimensional Wiener process \cite{Gar85} satisfying
\bqa
   {\rm E}[dW_{k}] &=& 0 \\
         dW_j(t)dW_k^*(t) &=& dt \, \delta_{jk},
   \label{ito1}\\
   dW_j(t)dW_k(t) &=& dt\, u_{jk} . \label{ito2}
\eqa
The only condition on the complex numbers $u_{jk}=u_{kj}$ is that
the corresponding complex symmetric matrix ${\bf u}$ must satisfy
\cite{WisDio01}
\beq
\Vert {\bf u} \Vert\leq 1.\label{unorm}
\eeq
This comes from the requirement that  the following expression
must be non-negative
\beq
    {\rm E}\sq{ \ro{\sum_{k} z_k dW_k + {\rm c.c}}^2}
    \label{mustbenn}
\eeq
for an arbitrary $K$-vector of complex numbers $\vec{z}$.

Some insight into the measurement parameters $u_{jk}$ may be found
by considering the simple case with one irreversible term; that
is, $K=1$ so that there is just one complex number $u$ in
\erf{ito2}. For specificity, say the system is an optical cavity
with annihilation operator $a$, damped through one end mirror with
decay rate $\kappa$. Then the continuous Markovian unravelings
correspond to two independent homodyne detection apparatuses
\cite{Car93b}, each of efficiency $1/2$. If the local oscillator
phases are $\theta_1$ and $\theta_2$ then $u=(e^{2i
\theta_1}+e^{2i \theta_2})/2$. The two photocurrents $I_{1}(t)$
and $I_{2}(t)$, normalized to have unit shot noise, are given by
\cite{Car93b,WisMil93a}
\beq
 I_{p}(t)dt = \sqrt{\kappa/2}\an{e^{-i\theta_{p}}a+e^{i\theta_{p}}a\dg}dt
+dW_{p}(t),
\eeq
where $dW_{1}$ and $dW_{2}$ are independent Wiener increments. We
can combine the photocurrents to make a complex signal
\bqa
J(t)dt &=&
[e^{i\theta_{1}}I_{1}(t)dt+e^{i\theta_{2}}I_{2}(t)dt]/\sqrt{2} \\
&=& \sqrt{\kappa}\an{a+ua\dg}dt+dW(t),
\eqa
where
$dW(t)=[e^{i\theta_{1}}dW_{1}(t)+e^{i\theta_{2}}dW_{2}(t)]/\sqrt{2}$
is a complex Wiener increment satisfying
\beq
dW^{*}(t)dW(t)=dt\;,\;\;dW(t)dW(t)=udt.
\eeq
That is, it has the same correlations as the $dW(t)$ occurring in the
stochastic master equation, and is in fact the same noise process.

If the two local oscillator phases are chosen to be identical then
$|u|=1$ and both apparatuses measure the same quadrature of the
cavity mode. If they are chosen to be in quadrature, with
$\theta_{1}-\theta_{2}=\pi/2$, then $u=0$ and two orthogonal
quadratures are measured each with efficiency $1/2$. In  general
$0\leq |u| \leq 1$, and for any $u\neq  0$, different amounts of
information are obtained about the two cavity-field quadratures.
The information gained tends to reduce the cavity field to a state
with correspondingly different quadrature uncertainties. This
gives an idea as to how different unravelings can give rise to
different ensembles.

For a master equation with $K$ Lindblad terms the problem of
finding the ensembles that are physically realizable by some
continuous Markovian unraveling (CMU) reduces to determining the
boundary $\{ u_{jk} \,:\, \Vert{\bf u}\Vert = 0 \}$ of a region in
$K(K+1)$-dimensional  Euclidean space. Even for a moderately
sized $K$ (for example $K=3$ is needed for the atom laser
problem), this is a surprisingly large space, which is difficult
to search efficiently. For that reason we adopt in this paper a
different search strategy, which will be explained in
Sec.~\ref{secss}.

\subsection{Quantum State Diffusion}

There is an interesting continuous Markovian unraveling, which has
some special properties, for the case where $u_{ij}\equiv 0$
\cite{Dio88,WisDio01}. In this case each complex Wiener process
$dW$ can be decomposed into real Wiener processes $dW^{a}, dW^{b}$
as
\beq
dW_{k}=(dW_{k}^{a}+idW_{k}^{b})/\sqrt{2}
\eeq
such that
$dW_{k}^{a}dW_{j}^{a}=dW_{k}^{b}dW_{j}^{b}=\delta_{jk}dt$, and
$dW_{k}^{a}dW_{j}^{b}=0$. This unraveling
is invariant under the complete set of linear transformations of
the Lindblad operators,
\beq
c_{\mu} \to U_{\mu\nu}c_{\nu}
\eeq
that leaves the master equation invariant. Here $U_{\mu\nu}$ is
an arbitrary unitary matrix.

This unraveling was introduced by Gisin and Percival
\cite{GisPer92}, under the name of quantum state diffusion (QSD),
as a microscopic model for decoherence. In the optical context, it
has been interpreted as the unraveling resulting from heterodyne
detection \cite{WisMil93c} or from equal-efficiency homodyne
detection of orthogonal quadratures (as discussed above), although
it can also arise in atomic detection schemes as well
\cite{VacRic98}. It has been suggested by Rigo and
Gisin\cite{RigGis96} that the QSD unraveling is a natural way to
discover the classical limit for a quantum system. Along similar
lines, Diosi and Kiefer \cite{DioKie00} have suggested that the
QSD unraveling is the most robust unraveling, or close to the most
robust unraveling (see the following paper \cite{WisVac01b} for a
detailed discussion of this concept). Thus, as well as considering
the set of all ensembles physically realizable from CMUs, we will
also pay particular attention to the ensemble arising from the
special instance of QSD.

\subsection{Discontinuous Unravelings} \label{discontun}

Although most of our calculations are restricted to CMUs, there
will be one occasion where we need to consider the following
discontinuous unravelings of the master equation (\ref{genme}):
\beq
   {\cal U}=\sum_{k}{\cal U}_{k},
\eeq
where
\beq
   {\cal U}_{k}(t)\rho dt
       = \left[dN_{k}(t) - dt\lambda_{k}(\rho)\right]
         \left( \frac{c_{k}\rho c_{k}\dg}{\lambda_k(\rho)}
              -\rho\right).
\eeq
Here the $dN_{k}(t)$ are point processes defined by
\bqa
   dN_{j}(t)dN_{k}(t) &=& \delta_{jk}dN_{k}(t) \\
   {\rm E}[dN_{k}] &=& \lambda_{k}(\rho)dt
          \,\equiv\, {\rm Tr}[\rho c_{k}\dg c_{k}]dt
\eqa
It is easy to verify that this unraveling satisfies the necessary
conditions of Eqs.~(\ref{prop1}) and (\ref{prop2}). This unraveling
(quantum jumps) is the most commonly used for numerical simulation of
master equations \cite{DumZolRit92}.

\section{The (Atom) Laser}

The system we wish to consider in this paper is the (atom) laser.
As noted in the introduction, we take a laser to be a device that
produces a coherent output, in the sense explained in
Ref.~\cite{Wis97}. An atom laser is thus a device that produces a
coherent beam of bosonic atoms, analogous to the coherent beam of
photons from an optical laser.

\subsection{The Master Equation}

A generic model for a laser was derived in Ref.~\cite{Wis97}. It
describes a single-mode field having annihilation operator $a$,
evolving under linear damping and nonlinear amplification. The
nonlinearity in the amplification is due to  depletion of the
source (the gain medium in optical lasers) and is essential for a
coherent output to form. In the interaction picture, and
measuring time in units of the decay rate, the master equation is
\beq
     \dot{\rho} = \mu  {\cal D}[a\dg]\left({\cal A}[a\dg]
              +n_{s}\right)^{-1}\rho + {\cal D}[a]\rho.
     \label{satlasme}
\eeq
The two terms on the right describe saturated gain and the decay
due to the coupling of the laser mode to the output beam,
respectively. Here $n_{s}$ is the saturation boson number, $\mu$
is a (typically) large parameter, ${\cal D}$ is as defined in
\erf{defcalD} and for arbitrary operators $A$ and $B$,
\beq
{\cal A}[A]B =  [A\dg A B + B A\dg A]/2.
\eeq

For simplicity we take the limit where $n_{s}$ can be ignored
compared to $a  a\dg$. Strictly this requires the limit $n_{s}
\ll 1$, because  the smallest eigenvalue of $a a\dg$ is 1.
However, for a laser at steady state the mean boson number is
typically much greater than 1, and only boson numbers close to
the mean are occupied with any significant probability. In the
above model the mean number is approximately $\mu - n_{s}$ in the
limit of large $\mu$. Hence in the limit $\mu \gg n_{s},1$ we can
ignore $n_{s}$ in \erf{satlasme}. The resultant
far-above-threshold laser master equation was first explicated
in Ref.~\cite{Wis93}.

Having made this simplification we now introduce more terms into
\erf{satlasme} in order to create a more realistic model. First,
we introduce a term describing phase diffusion. This will be
present in optical lasers for all sorts of technical reasons such
as thermal motion of the cavity mirrors. In an atom laser it may
also be present for more fundamental reasons, such as collisions
between uncondensed atoms (in the source modes) and atoms in the
laser mode condensate. Treating this phase diffusion as a
Markovian process, it is described by a Lindblad superoperator of
the form $N{\cal D}[a\dg a]$, where $N$ is the phase diffusion
rate in units of the decay rate.

The second new term we introduce is peculiar to atom lasers: the
self-energy of atoms in the condensate. This is described by a
Hamiltonian equal to $\hbar C(a\dg a)^{2}$, with
\beq
   C = \frac{2\pi\hbar a_{s} }{\kappa m}\int d^{3}{\bf r}|
          \psi({\bf r})|^{4},
   \label{selfenergy}
\eeq
where $\psi({\bf r})$ is the wavefunction for the condensate
mode, $a_{s}$ is the $s$-wave scattering length, and $\kappa$ is
the unit-valued decay rate of the condensate. Like the extra
phase diffusion term, this term has no effect on boson number; it
only affects the phase of the field. However it is strictly not a
phase diffusion term, but rather a dispersive term. It would
arise in an optical laser in a medium with a nonlinear
refractive index.

Putting the four terms (gain, loss, phase diffusion and
self-energy) together, the total master equation is
\bqa
\dot\rho &=& \left( \mu  {\cal D}[a\dg] {\cal A}[a\dg]^{-1}
 + {\cal D}[a] + N{\cal D}[a\dg a] \right)\rho \nn \\
&& -\, i C[(a\dg a)^{2},\rho]. \label{lasme}
\eqa
That this is of the Lindblad form follows from the identity
\beq
   {\cal D}[a\dg] {\cal A}[a\dg]^{-1}
        = \int_{0}^{\infty} dq {\cal D}[a\dg e^{-q a a\dg/2}].
   \label{linform}
\eeq
The stationary solution is a Poissonian mixture of number states
with mean $\mu$, just as expressed in Eqs.~(\ref{mixcoh}) and
(\ref{mixnum}):
\beq
   \rho_{\rm ss} = \int \frac{d\phi}{2\pi}\ket{\sqrt{\mu}
                    e^{i\phi}}\bra{\sqrt{\mu}e^{i\phi}}
                 = \sum_{n=0}^{\infty}e^{-\mu}\frac{\mu^{n}}{n!}
                     \ket{n}\bra{n} .
   \label{rhoinf}
\eeq

\subsection{The Linearized Master Equation}

The master equation (\ref{lasme}) is rather difficult to deal
with because of the nonlinearities in both the gain term and the
self-energy term. To make it more tractable we linearize this
equation for a state localized about a mean field $\langle a
\rangle = \sqrt\mu$. We make the replacement
\beq
a = \sqrt\mu +  (x + i y )/2
\eeq
and get, to second order in $x$ and $y$,
\bqa
\dot\rho &=& (1/4)\left\{ {\cal D}[ x + i
y] + (1+\nu) {\cal D}[ x]  + {\cal D}[ y]  \right. \nl{+} \left.
{\cal H}[i(xy+yx)/2-i\chi x^{2}] \right\} \rho, \label{linlasme}
\eqa
where
\beq \nu = 4N\mu \geq 0,\;\;
\chi = 4\mu C \label{nuchi}
\eeq
and ${\cal H}$ is the superoperator defined in \erf{defcalH},
which here is serving as a convenient way to describe the
Hamiltonian evolution. We have ignored a contribution to the
linearized Hamiltonian that is proportional to $a\dg a$ as this
simply indicates a frequency shift that can be removed in the
interaction picture.

To solve this master equation, we use the Wigner representation
$W(x,y)$ \cite{Gar91}. We make a Gaussian ansatz
\bqa
   W(x,y) &=& \exp\left[ \frac{\mu_{20}\mu_{02}}{\mu_{20}\mu_{02}-\mu_{11}^{2}}
              \left(- \frac{(x-\mu_{10})^{2}}{2\mu_{20}} \right.\right. \nn \\
          && \left.\left.\phantom{\times }
              +\,\frac{\mu_{11} (x-\mu_{10})(y-\mu_{01})}{\mu_{20}\mu_{02}}  -
               \frac{(y-\mu_{01})^{2}}{2\mu_{02}}\right)\right] \nn \\
           && \div \left( 2\pi \sqrt{\mu_{20}\mu_{02}-\mu_{11}^{2}} \right).
     \label{GW}
\eqa
Substituting this into \erf{linlasme} yields the following
ODEs for the moments
\bqa
\dot{\mu_{10}} &=& - \mu_{10} ,\\
\dot{\mu_{01}} &=& - \chi \mu_{10}, \\
\dot{\mu_{20}} &=& - 2\mu_{20}+2 ,\\
\dot{\mu_{11}} &=& - \mu_{11} - \chi\mu_{20}, \\
\dot{\mu_{02}} &=& - 2\chi\mu_{11} + 2+\nu .
\eqa
The solution is easy to find
\bqa
  \mu_{10}(t) &=& \mu_{10}(0) w ,
        \label{solns1}\\
  \mu_{01}(t) &=& \mu_{01}(0) - \chi \mu_{10}(0)(1-w),
        \label{meanph}\\
  \mu_{20}(t) &=& \mu_{20}(0) w^{2} + 1 - w^{2},
        \label{ampvar}\\
  \mu_{11}(t) &=& \mu_{11}(0) w - \chi\left\{ 1+w[\mu_{20}(0) - 2]
               \right.\nl\left.
               \phantom{\mu_{11}(0) w-\chi\left\{\right.}+\,
               w^{2}[1-\mu_{20}(0)] \right\} ,
        \label{covarsoln} \\
  \mu_{02}(t) &=& \mu_{02}(0) + (2+\nu)t -2\chi\mu_{11}(0)(1-w) \nl
          + \, 2\chi^{2}\left\{ t + [\mu_{20}(0) - 2] (1-w) \right.\nl\left.
          \phantom{2\chi^{2}\left\{\right.t}+\,  [1-\mu_{20}(0)] (1-w^{2})/2
          \right\}.
       \label{solns5}
\eqa
Here we are using the abbreviation $w\equiv e^{-t}$ .

\subsection{Coherence}  \label{seccoh}

Having solved for the dynamics of our (atom) laser model, we can
now answer the question, is it a true laser? That is, does it
satisfy the criteria for a coherent output as detailed in
Ref.~\cite{Wis97}. The first two criteria will be satisfied
provided the output coupling is realized in a suitable way. The
next two relate to the quantum noise of the state, and depend
upon the dynamics.

First, the laser intensity should be well-defined. Although this
criterion is strictly defined in terms of the output of the
laser, it will be satisfied if the boson number of the laser mode
itself is well-defined. In the present case this is clearly so
provided the mean number satisfies
\beq
\mu \gg 1,
\eeq
as the ratio of the standard deviation to the mean is equal to $1/\sqrt{\mu}$.

Second, the laser phase should be well-defined in the sense that
the phase should stay approximately constant over the time
between the emission of one boson and the next. With a unit
damping rate, this time is equal to $\mu^{-1}$. Rigorously, we
require that the magnitude of the first order coherence function
\beq
   g^{(1)}(t) =  {\an{a \dg(t) a(0)}}/{\an{a\dg a}}
   \label{g1}
\eeq
remain close to unity for $t=\mu^{-1}$. For the current system we
can rewrite this expression as
\bqa
  g^{(1)}(t)
      &=& \mu^{-1}{\rm Tr}[a\dg e^{{\cal L}t}(a\rho_{\rm ss})] \\
      &=& \mu^{-1}\int d\phi {\rm Tr}
          \left[a\dg e^{{\cal L}t}(a\ket{\sqrt{\mu}e^{i\phi}}
          \bra{\sqrt{\mu}e^{i\phi}})\right]
\eqa
Now because ${\cal L}$ is a phase-independent superoperator, the
trace here is independent of $\phi$. Thus the integral can be
dropped and we can rewrite this as
\beq
   g^{(1)}(t) = (1/\alpha^{*}){\rm Tr}\left[a \dg e^{{\cal L}t}
                      \ket{\alpha}\bra{\alpha}\right],
\eeq
where $|\alpha|^{2}=\mu$. Thus, the requirement that $g^{(1)}(t)\simeq
1$ for $t=\mu^{-1}$ is exactly  equivalent to
requiring that the system, initially in a coherent state of mean
number $\mu$, still have a phase variance much less  than
unity after a time $t=\mu^{-1}$.

Without loss of generality we can take the initial coherent state
to be $\ket{\sqrt{\mu}}$. Then
$\mu_{10}(0)=\mu_{01}(0)=\mu_{11}(0)=0$, $\mu_{20}=\mu_{02}=1$,
and $y$ is the phase quadrature. Assuming that the phase
uncertainty remains relatively small, we can make the
approximation
\beq
   \phi = \frac{y}{2\sqrt{\mu}}.
\eeq
From \erf{meanph}, the mean phase remains zero
\beq
   \an{\phi(t)} = \frac{\mu_{01}(t)}{2\sqrt{\mu}} = 0
\eeq
while the phase variance increases as
\beq
   \an{\phi^{2}(t)} = \frac{\mu_{02}(t)}{4\mu}.
\eeq

Substituting $t=\mu^{-1}\ll 1$ into \erf{solns5} yields
\beq
\an{\phi^{2}(t)} = \frac{1+ (2+\nu)\mu^{-1} +
\chi^{2}\mu^{-2}}{4\mu}.
\eeq
For the phase to remain well-defined we require this to be much
less than unity. Since we already require $\mu \gg 1$, this gives
the extra conditions
\bqa
\chi &\ll& \mu^{3/2}, \label{cohcon1}\\ \nu &\ll& \mu^{2}
\label{cohcon2}.
\eqa

In a typical optical laser (and certainly in some models of atom
lasers \cite{WisCol95}), $\nu \gg 1$. This means that excess
phase diffusion dominates the intrinsic phase diffusion (which
gives the $2$ in the $2+\nu$ term). In a typical atom laser, it
is also likely that excess phase diffusion will dominate.
However, as long as $\nu \ll \mu^{2}$ the laser will remain
coherent. Since $\nu=4N\mu$, this is equivalent to the condition
\beq
N \ll \mu.
\eeq
This expression places an upper bound on the phase diffusion rate
$N$ for the device to be considered a laser.

For an optical laser any nonlinear refractive index is usually
small and $\chi \ll 1$. For an atom laser $\chi$ is likely to be
much greater than one, as we will discuss in Sec.~\ref{sec:expt}.
To be a true atom laser it is necessary for
it to remain much less than $\mu^{3/2}$. Since $\chi=4\mu C$ the
phase coherence condition  places an upper bound on the condensate
self-energy in \erf{selfenergy} of
\beq
C \ll \mu^{1/2}.
\eeq

\section{Unraveling the (Atom) Laser}  \label{secMRU}

We now wish to consider monitoring the environment of the laser in
order to realize physically an ensemble of pure states. This would
be very to difficult to do experimentally, as it would require
monitoring all reservoirs for the device, including the source of
bosons (the gain medium) and the sources of phase diffusion as
well as the laser output. However in principal these things can be
done providing the laser evolution is well-approximated by a
Markovian master equation.

\subsection{Realizing the Number State Ensemble} \label{secrnse}

Before turning to continuous Markovian unravelings, we consider a
discontinuous unraveling to show how the ensemble consisting of
number states can always be realized. Using the atom laser master
equation  (\ref{lasme}) in the Lindblad form (\ref{linform}), we
can apply the unraveling of Sec~\ref{discontun}, where the
Lindblad operators are
\bqa
c_{0} &=& a \\
c_{N} &=& \sqrt{N}a\dg a ,
\eqa
plus a continuum of Lindblad operators
\beq
c_{q} = \sqrt{\mu}a\dg e^{-q a a\dg/2}, \textrm{ for } q\in[0,\infty).
\eeq
Each of these operators either leaves the a number state
unchanged, or turns it into another number state. Since the
Hamiltonian $Ca\dg a\dg aa$ also leaves a number state unchanged, it
follows that if the system is initially in a number state,
it will simply jump between number states under this unraveling.
Moreover, it can be shown that an arbitrary initial state will tend
towards some number state under this unraveling. In this way it is
clear that the number state ensemble (\ref{mixnum}) can always be
physically realized.

\subsection{The Continuous Markovian Unravelings}

As mentioned in Sec.~\ref{seccmu},  we
are principally concerned with continuous Markovian unravelings. In
this case, from
the master equation (\ref{lasme}), the SME  is
\bqa
d{P} &=& dt\left\{ \mu \int_{0}^{\infty} dq {\cal D}[a\dg e^{-q a
a\dg /2}] +{\cal D}[a] + N{\cal D}[a\dg a] \right\}P \nl  +\,
\sqrt{\mu} \int_{0}^{\infty} dq {\cal H}[dW_{q}^{*}(t)a\dg e^{-q
a a\dg /2}]P   \nl +\,{\cal H}[dW_{0}^{*}(t)a]P+\sqrt{N} {\cal
H}[dW_{N}^{*}(t)a\dg a] P \nl -\, i dt [C(a\dg a)^{2},P].
\label{tobelin}
\eqa
Here $dW_{0}$ is a zero-mean white
noise term. If we define $\zeta_{0}(t)=dW_{0}(t)/dt$ we have
\beq
{\rm E}[\zeta^{*}_{0}(t)\zeta_{0}(t')] = \delta(t-t').
\eeq
and likewise for $\zeta_{N}$ and $\zeta_{q}$ for each $q$. We say that
these white noise terms are {\em distinct} because the
cross terms are zero, for example
\beq
{\rm E}[\zeta^{*}_{0}(t)\zeta_{N}(t')] = 0.
\eeq

Now we wish to linearize. First note that

\singlecol

\bqa
\sqrt\mu\int_{0}^{\infty} dq \zeta_{q}^{*}(t)a\dg\exp(-q a a\dg /2)
&\simeq& \sqrt\mu\int_{0}^{\infty} dq \zeta_{q}^{*}(t) e^{-\mu
q/2}\sqrt{\mu}
\left[
1 +  (x + iy - \mu xq)/2\sqrt{\mu}\right] \\
&=& c{\rm -number} + \frac{y}{2} \int_{0}^{\infty} dq\,
i\zeta_{q}^{*}(t) e^{-\mu q/2}
 + \frac{x}{2} \int_{0}^{\infty} dq\, \zeta_{q}^{*}(t) e^{-\mu q/2}(1-q) \\
&\equiv& c{\rm -number} + \left[y \zeta_{2}^{*}(t) +  x
\zeta_{3}^{*}(t)\right]/2, \label{SMElin}
\eqa
where $\zeta_{2}(t)$ and $\zeta_{3}(t)$ are distinct
complex normalized white noise terms
as usual.

Using this, we linearize \erf{tobelin} as
\bqa
d{P} &=& (1/4) dt\left\{ {\cal D}[ x + iy]\rho +(1+\nu) {\cal D}[
x]\rho + {\cal D}[ y] \rho + {\cal H}[i (xy+yx)/2] +  {\cal
H}[-i\chi x^{2}] \right\}P \nn \\ && +\, (1/2) \left\{ {\cal
H}[dW_{0}^{*}(t)  (x+iy)]  + \sqrt{1+\nu}\, {\cal H}[dW_{1}^{*}(t)  x]
+ {\cal H}[dW_{2}^{*}(t)  y] \right\} P.
\label{lassse}
\eqa

\doublecol

where we have defined a new white noise source $\sqrt{1+\nu} \,
dW_{1}(t) = dW_{3}(t) + \sqrt{\nu}\,dW_{N}(t)$. We could have
obtained this result directly from the linearized form of the
master equation (\ref{linlasme}), but this derivation makes the
physical origin of the noise terms apparent.

The three complex white noise sources $dW_{j}=\zeta_{j}dt$ are
distinct in the above sense that
\beq
{\rm E}[\zeta_{i}^{*}(t) \zeta_{j}(t')] = \delta_{ij}\delta(t-t').
\eeq
However they
can still be correlated in the sense that
\beq
   {\rm E}[\zeta_{i}(t) \zeta_{j}(t')] = u_{ij}\delta(t-t'),
   \label{corr_coeff}
\eeq
where the $u_{ij}$ are constrained only by \erf{unorm}.
The $\delta$-function in time in
\erf{corr_coeff} is not required to reproduce the master
equation. It is a consequence of our restriction to Markovian
unravelings.

Now it is a remarkable fact about the stochastic master equation
(\ref{lassse}) that it takes Gaussian states to Gaussian states.
This will be true for any diffusive stochastic master equation
that is at most second-order in $x$ or $y$.
The significance in this case is that we can again
use the ansatz (\ref{GW}), and we need only the equations of
motion for the five moments. We find the following equations (to
be interpreted in the \ito sense \cite{Gar85})
\singlecol

\bqa
d\mu_{10}/dt &=& -\mu_{10} + {\rm Re}\left\{ \zeta_{0}^{*}(t)  \left[
\mu_{20} -1 + i\mu_{11}\right] +
  \zeta_{1}^{*}(t) \sqrt{1+\nu} \left[
\mu_{20}\right] +
  \zeta_{2}^{*}(t) \left[
\mu_{11}+i\right] \right\} \label{first}\\
d\mu_{01}/dt &=& -\chi \mu_{10} + {\rm Re}\left\{ \zeta_{0}^{*}(t)
\left[i\mu_{02} -i + \mu_{11}\right] +
 \zeta_{1}^{*}(t) \sqrt{1+\nu}\left[
\mu_{11}-i \right]  +
 \zeta_{2}^{*}(t) \left[
\mu_{02}\right] \right\} \label{second}\\
d\mu_{20}/dt &=& 2-2\mu_{20} -  {\rm Re}\left[
(\mu_{20}-1)^{2}+ \mu_{11}^{2}+ (1+\nu)\mu_{20}^{2}+\mu_{11}^{2}+1
\right. \nn \\
&& +
u_{00}^{*}(\mu_{20}-1+i\mu_{11})^{2} + u_{11}^{*}(1+\nu) \mu_{20}^{2}
 + u_{22}^{*}(\mu_{11}+i)^{2} \nn \\
&& \left. +\, 2u_{01}^{*}\sqrt{1+\nu}\,
(\mu_{20}-1+i\mu_{11})\mu_{20} +
2u_{02}^{*}(\mu_{20}-1+i\mu_{11})(\mu_{11}+i) +
2u_{12}^{*}\sqrt{1+\nu}\,(\mu_{11}+i)\mu_{20} \right]/2
\label{mu20}\\ d\mu_{02}/dt &=& -2\chi \mu_{11}+2+\nu -  {\rm
Re}\left[ (\mu_{02}-1)^{2}+ \mu_{11}^{2}+ (1+\nu)(\mu_{11}^{2}+1)
+
\mu_{02}^{2} \right. \nn\\
&&+ u_{00}^{*}(i\mu_{02} -i + \mu_{11})^{2}
+u_{11}^{*}(1+\nu)(\mu_{11}-i )^{2} + u_{22}^{*}\mu_{02}^{2}
\nn\\ && \left. +\, 2u_{01}^{*}\sqrt{1+\nu}\,
(i\mu_{02}-i+\mu_{11})(\mu_{11}-i) +
2u_{02}^{*}(i\mu_{02}-i+\mu_{11})\mu_{02} +
2u_{12}^{*}\sqrt{1+\nu}\,(\mu_{11}-i)\mu_{02} \right]/2
\label{mu02}\\
d\mu_{11}/dt &=& -\mu_{11} - \chi \mu_{20} - {\rm
Re}\left\{ (\mu_{20} -1 + i\mu_{11})(-i\mu_{02} +i + \mu_{11}) +
(1+\nu)(\mu_{11}-i)\mu_{20} +
\mu_{02}(\mu_{11}-i) \right. \nn \\
&& + u_{00}^{*}(\mu_{20} -1 + i\mu_{11})(i\mu_{02} -i + \mu_{11})
+ u_{11}^{*}(1+\nu) \mu_{20}(\mu_{11}-i)
 + u_{22}^{*}\mu_{02}(\mu_{11}+i) \nn \\
&& +\, u_{01}^{*}\sqrt{1+\nu}\,
[(\mu_{20} -1 + i\mu_{11})(\mu_{11}-i) + \mu_{20}(i\mu_{02} -i +
\mu_{11})]  +
u_{12}^{*}\sqrt{1+\nu}\,[\mu_{20}\mu_{02} +
(\mu_{11}+i)(\mu_{11}-i)]  \nn\\
&& +\left. u_{02}^{*}[(i\mu_{02}-i+\mu_{11})(\mu_{11}+i) +
(\mu_{20}-1+i\mu_{11})\mu_{02}] \right\}/2
\label{mu11}
\eqa

\doublecol

\subsection{The Stationary Solutions}
\label{secss}

From these equations we see that the evolution of the second order
moments $\mu_{20},\mu_{02},\mu_{11}$ is deterministic. This means
that for a given unraveling ${\cal U}$ the stationary ensemble
will consist of Gaussian pure states all having the same second
order moments. They are distinguished only by their first order
moments $\bar{x}=\mu_{10},\bar{y}=\mu_{01}$, which therefore take
the role of the index $n$ in \erf{ens}. The different ensembles
themselves are indexed by another pair of numbers,
$\mu_{11},\mu_{20}$, which play the role of ${\cal U}$ in
\erf{ens}. We do not need $\mu_{02}$ because the purity of the
unraveled states implies that
\beq
   \mu_{20}\mu_{02} - \mu_{11}^{2} = 1. \label{pur}
\eeq
However, it should be noted that the mapping from ${\cal U}$ to
$\mu_{11},\mu_{20}$ is in general many-to-one as discussed below.

We now introduce a new notation for the second order moments,
\beq
\alpha = \mu_{02}\;;\;\;\beta = \mu_{11}\;;\;\; \gamma = \mu_{20},
\eeq
The different ensembles are now indexed by the pair
$\beta,\gamma$. Of course not all pairs $\beta,\gamma$ correspond to physically
realizable ensembles. Since the ensemble we are considering has
evolved to a steady state at $t=0$, the only valid pairs must
satisfy Eqs.~(\ref{mu20})--(\ref{mu11}) with the left-hand sides
set to zero. This gives three simultaneous equations that, on
splitting  $u_{ij}$ into real $r_{ij}$ and imaginary $h_{ij}$
components, can be written as

\singlecol

\bqa
  1-\gamma-(1+\nu/2)\gamma^{2}-\beta^{2} &=&
        r_{00}[(\gamma-1)^{2}-\beta^{2}]/2 + r_{11}(1+\nu)\gamma^{2} /2
        + r_{22} (\beta^{2}-1) /2\nn \\
     &&+\,h_{00} \beta(\gamma-1) +h_{22}\beta \nn\\
     &&+\,r_{01}\sqrt{1+\nu}\,\gamma(\gamma-1)+
        r_{02}(\gamma-2)\beta+r_{12}\sqrt{1+\nu}\,\gamma\beta\nn \\
     &&+\, h_{01}\sqrt{1+\nu}\,\gamma\beta
       + h_{02}(\beta^{2}+\gamma-1) + h_{12}\sqrt{1+\nu}\,\gamma
     \label{lin_sys1}\\
  -2\chi\beta + (1+ \nu/2)(1-\beta^{2})-\alpha^{2}+\alpha
     &=& r_{00}[\beta^{2}-(\alpha-1)^{2}]/2 +
         r_{11}(1+\nu)(\beta^{2}-1)/2 + r_{22}\alpha^{2}/2 \nn \\
      &&+\,h_{00}\beta(\alpha-1) + h_{11}(1+\nu)(-\beta) \nn \\
      &&+\,r_{01}\sqrt{1+\nu}\,(\beta^{2}+\alpha-1) + r_{02}\beta\alpha +
         r_{12}\sqrt{1+\nu}\,\beta\alpha \nn\\
      &&+\,h_{01}\sqrt{1+\nu}\,(\alpha-2)\beta + h_{02}(\alpha-1)\alpha
         + h_{12}\sqrt{1+\nu}\,(-\alpha)
      \label{lin_sys2}\\
  -\chi\gamma -\alpha \beta -(1+\nu/2)\gamma\beta
     &=& r_{00}\beta(\gamma-\alpha)/2 + r_{11}(1+\nu)\gamma\beta/2
          + r_{22}\beta\alpha/2 \nn \\
     && +\, h_{00}[\beta^{2}+(\alpha-1)(\gamma-1)]/2 +
          h_{11}(1+\nu)(-\gamma)/2 + h_{22}\alpha/2 \nn \\
     && +\, r_{01}\sqrt{1+\nu}\,\gamma\beta +
        r_{02}[\beta^{2}+ 1 + (\gamma - 2)\alpha]/2+ 
        + r_{12}\sqrt{1+\nu}\,(\alpha\gamma+\beta^{2}+1)/2 \nn \\
     && +\,h_{01}\sqrt{1+\nu}\,[\beta^{2}+1+(\alpha-2)\gamma] /2 +
         h_{02}\beta\alpha,     \label{lin_sys3}
\eqa \doublecol
where $\alpha$ is to be read as $(1+\beta^{2})/\gamma$.

These three equations are nonlinear in $\beta,\gamma$ but linear
in the 12 real variables $r_{ij},h_{ij}$. This means that if the
values of $\gamma$ and $\beta$ are known then the three equations
can be solved for $r_{ij}, h_{ij}$.  Since there are only three
equations for the 12 unknown variables, the resulting linear
system is non-singular and an (uncountably) infinite number of
solutions are possible. We denote the family of such solutions
$F_\xi=\{r^{(\xi)}_{ij},h^{(\xi)}_{ij}:i,j=0,1,2\}$,  indexed by
$\xi$. Physically this arises because many different unravelings
${\cal U}$ may lead to the same steady state ensemble
$\beta,\gamma$. The question of whether a given pair of values of
$\gamma$ and $\beta$ represents a physically realizable state then
becomes the problem of determining whether any of the solutions
$F_\xi$ for the correlation coefficients
$u^{(\xi)}_{ij}=r^{(\xi)}_{ij}+ih^{(\xi)}_{ij}$ satisfy the
condition $\Vert {\bf u}^{(\xi)}\Vert \leq 1$ in \erf{unorm}. This
problem can be solved by finding the solution $F_{\Xi}$ that gives
the smallest value for $\Vert {\bf u}\Vert$, and checking if this
is less than one.

The above method determines the boundary between those ensembles
that are physically realizable and those that are not by finding,
explicitly, the parameters of the unravelings that satisfy $\Vert
{\bf u}^{(\xi)}\Vert = 1$. There is an alternate, but equivalent,
approach \cite{WisVac01c} based on the central idea of
Ref.~\cite{HugJozWoo93}. This allows one to take an arbitrary
ensemble and check whether it is possible, by monitoring the
environment, for the state of the system to be restricted to
members of the ensemble over arbitrary time intervals. The
ensemble is physically realizable if, and only if, this can be
done without changing the ensemble average dynamics. The advantage
of this alternate approach is that the parameters of the
unraveling need not be calculated explicitly and so the
computational task can be greatly reduced. Moreover it is possible
to find the boundary between physically realizable and non
realizable ensembles in a closed analytic form. The details are
tangential to the scope of the present paper and are explored
elsewhere \cite{WisVac01c}. We note here however that the PR
region is given by $\beta$ and $\gamma$ values satisfying $\gamma
> 0$ and
\bqa
  (2\chi\beta-2-\nu)(2-2\gamma)-(\beta+\chi\gamma)^2 \ge 0.
  \label{PRboundary}
\eqa
We have verified this analytic result with numerical solutions
obtained using the former approach, for all cases presented below.

\subsection{The Stationary Ensemble}

The stationary solution of the linearized
master equation (\ref{linlasme}) has a Wigner function
which is independent of phase ($y$) and has the following
amplitude
($x$) dependence:
\beq
   W_{\rm ss}(x) \propto (2\pi)^{-1/2}\exp( - x^{2}/2).
   \label{Wss}
\eeq
This is as expected from the stationary solution of the full
master equation, \erf{rhoinf}. A flat phase distribution
linearizes into a flat $y$-distribution.

As shown above, the long-time solution of the SME (\ref{SMElin})
is an ensemble of Gaussian pure states in which the second order
moments $\mu_{20},\mu_{11},\mu_{02}$ are identical in all members
of the ensemble, but $\bar{x}=\mu_{10}$ and $\bar{y}=\mu_{01}$
are allowed to vary. The ensemble is thus represented as
\beq
   E^{\cal U} = \{(\wp^{\cal U}_{\bar{x},\bar{y}},P^{\cal
   U}_{\bar{x},\bar{y}}):{\bar{x},\bar{y}}\in \Re\},
   \label{EcalU}
\eeq
where the second order moments of the pure state
$P^{\cal U}_{\bar{x},\bar{y}}$ are determined by the unraveling ${\cal U}$.

The weighting function $\wp_{\bar{x},\bar{y}}^{\cal U}$ for the
members of the ensemble is Gaussian. This follows from the fact
that Eqs.~(\ref{first}), (\ref{second})  for $\bar{x}$ and
$\bar{y}$ describe in steady state (where the second-order
moments are constant) a two-dimensional Ornstein-Uhlenbeck
process \cite{Gar85}. Such a process has a stationary probability
distribution that is Gaussian.

Rather than deriving this stationary Gaussian distribution
$\wp_{\bar{x},\bar{y}}^{\cal U}$ from the Ornstein-Uhlenbeck
process we can derive it more simply by noting that it must
satisfy
\beq
   \rho_{\rm ss} = \int d\bar{x} \,d\bar{y}\,
   \wp^{\cal   U}_{\bar{x},\bar{y}}P^{\cal U}_{\bar{x},\bar{y}}.
\eeq
This is guaranteed by the fact that the SME is equivalent to the
master equation on average. Evidently  $\bar{y}$ should always
have a flat weighting distribution, and $\bar{x}$ should have the
weighting distribution
\beq
   \wp^{\cal   U}(\bar{x}) = [2\pi(1-\mu_{20})]^{-1/2}
   \exp\left[-  \bar{x}^{2} / 2(1-\mu_{20})\right].
   \label{Pbarx}
\eeq
This ensures that
\beq
W_{\rm ss}(x) \propto \int d\bar{y}  \int d\bar{x}\, \wp^{\cal
U}(\bar{x})
W^{\cal U}_{\bar{x},\bar{y}}(x,y),
\eeq
where $W^{\cal U}_{\bar{x},\bar{y}}(x,y)$ is the Wigner function of
$P^{\cal U}_{\bar{x},\bar{y}}$.

\section{PR Ensembles for the (Atom) Laser}

In this section we present our results for the physically
realizable ensembles for the (atom) laser.

\subsection{Realizing the Number State Ensemble}

Before turning to the effect of varying the dynamical parameters
$\chi$ and $\nu$ we briefly return to the physical realizability
of the number state ensemble. We showed in Sec.~\ref{secrnse}
above that this ensemble can be realized by a discontinuous
unraveling. The analog of the number states in the linearized
regime we have been considering are the infinitely squeezed states
with $\gamma=\beta=0$, $\alpha=\infty$. We expect that these
states should be PR using a CM unraveling. This expectation is
met, in that these state parameters are a solution of
Eqs.~(\ref{lin_sys1})--(\ref{lin_sys3}) for $u_{00}=1$,
$u_{11}=1$, $u_{22}=-1$, and all other $u_{jk}=0$.

\subsection{Varying $\chi$ with $\nu=0$}

\label{svarchi}
First we present the results showing the effect of varying $\chi$
for fixed $\nu=0$. As we have established above, a PR ensemble
from a CMU can be represented by the pair of numbers
$\gamma,\beta$. Thus the set of all PR ensembles can be
represented by a region in $\gamma-\beta$ space
$[0,1]\times(-\infty,\infty)$. The boundaries of this region,
given by \erf{PRboundary}, are shown in Fig.~\ref{fig1} for
various values of $\chi$. A number of features of this plot are
evident. First, for any non-zero value of $\chi$, the coherent
state ensemble is not PR. Second, as $\chi$ increases the PR
ensembles become increasingly removed from the coherent state
ensemble. Third, the boundary of the PR ensembles is asymmetric in
$\beta$ for $\chi>0$, with a larger negative $\beta$ region.

The first point can easily be proven analytically. Coherent states are
given by $\alpha=\gamma=1$ and $\beta=0$ for which
\erf{PRboundary} gives $-\chi^2\ge 0$.  That is, coherent states
are physically realizable only for $\chi=0$.

We quantify the second point by defining the {\em
closest-to-coherent (CC) ensemble} as that for which the states
have maximum overlap with a coherent state. The overlap of two
Gaussian states with the same mean amplitudes and covariance
parameters $\alpha,\beta,\gamma$ and
$\alpha_{1},\beta_{1},\gamma_{1}$ is
\beq
   2/\sqrt{(\alpha_{1}+\alpha)(\gamma_{1}+\gamma)-
   (\beta_{1}+\beta)^{2}}
\eeq
If one of these is a coherent state, with
$\alpha_{1}=\gamma_{1}=1$, $\beta_{1}=0$, this reduces to
\beq
   2/\sqrt{ 2 + \alpha + \gamma}.
\eeq
Thus, to find the closest-to-coherent ensemble we simply find the
minimum  $\alpha+\gamma = \gamma+(1+\beta^{2})/\gamma$ in the PR
region of $\gamma-\beta$ space.

The closest-to-coherent ensemble for each value of $\chi$ is
represented in Fig.~\ref{fig1} as a filled circle on the boundary
of the respective PR region.  The states in these ensembles become
more squeezed ($\gamma\to 0$) and have a greater $x-y$ covariance
as $\chi$ increases. This trend is shown in more detail in
Fig.~\ref{fig2} where we plot the parameters $\alpha$, $\beta$ and
$\gamma$ for the closest-to-coherent PRE as a function of $\chi$.
By finding the minimum of $\gamma+(1+\beta^{2})/\gamma$ subject to
the constraint \erf{PRboundary} and expanding about $\gamma=0$ and
$1/\chi=0$ we find the parameters of the CC ensemble for large
$\chi$ scale as
\bqa
  \alpha^{\rm CC} &\simeq&  \frac{2}{3^{3/4}} \chi^{1/2}, \label{alphaCC}\\
  \gamma^{\rm CC} &\simeq&  \frac{2}{3^{1/4}}  \chi^{-1/2}, \label{gammaCC}\\
  \beta^{\rm CC}  &\simeq&  -\frac{1}{3^{1/2}}.\label{betaCC}
\eqa
Also plotted in the figure are two lines representing $\chi^{1/2}$
and $\chi^{-1/2}$ for comparison.  One can clearly see the $1/2$
power law scaling for $\alpha$ and $\gamma$.

The third point, i.e. the increasing asymmetry of the PR regions
in Fig.~\ref{fig1}, is due to the self-energy of the condensate
embodied by the term containing $(a^\dagger a)^2$ in \erf{lasme}.
In the Wigner phase-space representation, this term by itself
produces a `phase shearing'; that is, the angular velocity of the
point $(x,y)$ depends on the distance $\simeq \sqrt{\mu}(1+x)$ from the
origin \cite{anharmonic_milburn}.  In our linearized model of the
atom laser, the effect of this term is to shear the circular
contours of a coherent state into ellipses.  \erf{covarsoln}
indicates that these ellipses have a negative covariance.  When
monitoring the reservoirs it will therefore be easier to realize
states with a negative covariance. Hence, the PR regions become
more asymmetric allowing more negative-$\beta$ regions as the
nonlinearity parameter $\chi$ increases.

\subsection{The effect of nonzero $\nu$}

Non-zero values of $\nu$, as defined in \erf{nuchi}, correspond to
the presence of excess phase diffusion, which will tend to overcome
the phase-shearing effect. This makes it easier to physically
realize states that are closer to coherent states.  In
Fig.~\ref{fig3} we plot the boundaries of the PR ensembles for
$\nu=10$ for the same set of values of $\chi$ as in
Fig.~\ref{fig1}.  The CC ensembles are also shown as filled
circles. The PR regions are generally broader as expected, and
this allows the CC ensemble to be closer to a coherent state than
for corresponding $\chi$ values in Fig.~\ref{fig1}.

The parameters for the CC ensembles for $\nu=100$ as a function of
$\chi$ are plotted in Fig.~\ref{fig4}. Comparing with
Fig.~\ref{fig2} we note that the presence of the excess phase
diffusion in Fig.~\ref{fig4} allows ensembles very close to
coherent states (i.e. with $\alpha\approx\gamma\approx 1$,
$\beta\approx 0$) for $\chi$ up to of order $\nu^{1/2}$. This can be
verified analytically from \erf{PRboundary}.
 However, as the value
of $\chi$ increases beyond this to of order $\nu$,
the effect of the non-zero $\nu$ value
becomes less significant and the curves approach the same
asymptotes as in Fig.~\ref{fig2}.

The physically-realizable region for $\chi=0$ includes the point
$\beta=0$, $\gamma=1$ for all values of $\nu$. Hence the
closest-to-coherent PR ensemble is trivially an ensemble of
coherent states in this case. The situation is different for
nonzero $\chi$. Fig.~\ref{fig5} shows the parameters for the
closest-to-coherent PR ensemble as a function of $\nu$ for
$\chi=100$.  For $\nu\approx 0$ the values of $\alpha$, $\beta$
and $\gamma$ are approximately the same as the corresponding
values at $\chi=100$ in Fig.~\ref{fig2}. However, as $\nu$
increases much larger than $\chi$, the effect of the self-energy term
become less significant and the phase diffusion begins to dominate.
Then, for $\nu \agt \chi^{2}$ the closest-to-coherent ensemble approaches a set
of coherent states as $\alpha, \gamma \to 1$.

\subsection{Comparison with quantum state diffusion}

The unraveling given by quantum state diffusion (QSD) is more
restrictive than that of the general continuous Markovian
unraveling treated here.  Specifically, for the QSD unraveling, $\alpha$,
$\beta$ and $\gamma$ must satisfy
Eqs.~(\ref{lin_sys1})-(\ref{lin_sys3}) for
$u_{ij}=r_{ij}+ih_{ij}=0$ instead of any $u_{ij}$ fulfilling
$\Vert {\bf u}\Vert \leq 1$. We find this yields the analytic
solutions for the QSD ensemble
\bqa
  \alpha^{\rm QSD}  &=& \frac{1+\sqrt{1 - 8\chi\beta + 4M(1-\beta^2)}}{2}
        \label{alphaQSDan}\\
  \gamma^{\rm QSD}  &=& \frac{-1 + \sqrt{1 + 4M(1 - \beta^2)}}{2M}\\
  \beta^{\rm QSD}  &=& \frac{(-1 + 4M - F)\chi + \sqrt{G-E}}
             {4(\chi^2 + M)}
        \label{betaQSDan}
\eqa
where
\bqa
   M &\equiv& 1 + \nu/2\\
   E &\equiv& (24M - 2)\chi^2 + 32M^3 + 8M^2\\
   F &\equiv& 4 \sqrt{(M + {1}/{4})^2 + \chi^2}\\
   G &\equiv& 2(4M^2 + \chi^2)F\ .
\eqa
The crosses in Figs.~\ref{fig1} and \ref{fig3} represent the QSD
ensembles for the same set of $\chi$ and $\nu$ values as the CC PR
ensembles. The corresponding value of $\chi$ for the crosses
reduces from left to right.  One immediately notices that the QSD
ensembles lie well inside of the PR boundary indicating that, for
moderate $\chi$ and $\nu$ values, the QSD unraveling is
significantly more restrictive than the general continuous
Markovian unraveling explored here. Moreover, the QSD ensembles
are more squeezed (smaller $\gamma$ values) than the corresponding
CC ensembles.

We note that the QSD ensemble is significantly squeezed even for
the ideal photon laser limit of $\chi=\nu=0$ for which the QSD
ensemble is given by $\alpha=(\sqrt{5}+1)/2\approx 1.62$,
$\beta=0$ and $\gamma=(\sqrt{5}-1)/2\approx 0.62$. We can trace
the origin of this squeezing as follows.  The second term on the
right side of \erf{lasme} represents the output coupling of the
laser. As mentioned above, QSD corresponds to equal-efficiency
homodyne detection of a pair of orthogonal quadratures.  Thus, in
QSD the monitoring of the output will tend to localize the state
of the laser onto a coherent state. No squeezing can therefore
originate from this term.  The squeezing must therefore originate
from the nonlinear amplification process represented by the first
term on the right side of \erf{lasme}. Indeed, the nonlinear
amplification restricts the amplitude noise through depletion of
the source.  In our linearized model, this corresponds to
restricted noise in $x$. Evidently, the monitoring of the
reservoir modes associated with the amplification is a partial
measurement of $x$ and this leads to the squeezing of $x$.

It is interesting to compare this with the general continuous
Markovian unraveling (CMU) treated in the previous subsection.
This is less restrictive than QSD since, for example, it allows
the {\em unbalanced} monitoring of two quadratures of the output
field. In particular, a correlation value of $u_{00}=-1$
corresponds to the monitoring of just the $y$ quadrature. This
would tend to localize the state of the laser mode onto a state
with reduced $y$ fluctuations and thus counteract the
$x$-quadrature squeezing effect from the nonlinear amplification.
Similar remarks apply to unraveling the gain process itself. The
net effect is that the general continuous Markovian unravelings
can physically realize coherent states for $\chi=\nu=0$ whereas
QSD does not.

Despite these differences,  the $\alpha$ and $\gamma$ scaling laws
for the QSD ensemble follow the same $\chi^{\pm 1/2}$ power laws
as the closest-to-coherent ensemble although with a different
prefactor. In Fig.~\ref{fig6} we plot the parameters for the QSD
ensemble for $\nu=0$ as a function of $\chi$. Comparing with
Fig.~\ref{fig2} we note that the QSD ensemble begins more squeezed
for small $\chi$, but for large $\chi$ the two ensembles approach
similar degrees of squeezing.  In fact, from
Eqs.~(\ref{alphaQSDan})-(\ref{betaQSDan}) we find the scaling laws
\bqa
  \alpha^{\rm QSD} &\simeq&  \sqrt{2} \chi^{1/2}, \\
  \gamma^{\rm QSD} &\simeq& \sqrt{2} \chi^{-1/2}, \\
  \beta^{\rm QSD} &\simeq&  -1
\eqa
which should be compared with Eqs.~(\ref{alphaCC})-(\ref{betaCC}).

In Fig.~\ref{fig7} we plot the parameters for QSD ensemble for
$\chi=0$ as a function of $\nu$.  The QSD ensembles are highly
squeezed for increasing $\nu$ and, indeed, we find
\bqa
  \alpha^{\rm QSD} &\simeq&  \frac{1}{\sqrt{2}} \nu^{1/2}, \\
  \gamma^{\rm QSD} &\simeq& \sqrt{2} \nu^{-1/2}, \\
  \beta^{\rm QSD} &=&  0 \ .
\eqa
This is perhaps surprising given that one does not usually
associate enhanced squeezing with large phase diffusion. However,
the monitoring of the reservoir corresponding to the phase
diffusion is effectively an incomplete measurement of the variable
$a^\dagger a$, which, in our linearized model \erf{linlasme}, is
represented by the term $\nu {\cal D}[x/2]$. The monitoring therefore tends
to localize the state of the laser onto an eigenstate of $x$.  The
strength or rate of these measurements increases with $\nu$. In
QSD there is no mechanism to counteract the associated squeezing
of the $x$-quadrature, and so the squeezing increases with $\nu$.
In contrast, the general continuous Markovian unraveling allows
unbalanced monitoring of all baths. In particular, with $u_{11}=-1$,
the phase diffusion is  unraveled as a pure noise process
(stochastically changing the phase of the state, but yielding no
information about it).  This allows the closest-to-coherent CMU ensemble to
comprise of coherent states for the same parameters as for
Fig.~\ref{fig7}.

\section{Discussion}

\subsection{Summary}

The atom laser, even under with the simplifying approximations we
have made, is an open quantum system with rich dynamics. Some
aspects of the dynamics, such as excess phase diffusion
(parametrized by $\nu$) and phase dispersion caused by atomic
interactions (parametrized by $\chi$), do not affect the
stationary state. That is because the stationary state is a
Poissonian mixture of number states. In this paper we have
investigated the representations of this mixed state as ensembles
of pure states. The diagonal representation (number states) is
one such ensemble, and the random-phase coherent state ensemble
is another. Although mathematically equivalent we have found that
such representations are not physically equivalent, as only some
of them can be physically realized through monitoring the system.
Moreover, the dynamical parameter $\chi$, which does not affect
the stationary state at all, radically affects which pure state
ensembles are physically realizable (PR). In particular, for any
$\chi \neq 0$, the ensemble of coherent states with unknown phase
is not PR.

As the nonlinearity $\chi$ is increased, the PR ensembles become
increasingly removed from the coherent state ensemble. To be
specific, the ensemble of states that are closest to coherent
states consists of states that are amplitude squeezed (but
slightly rotated), with a phase quadrature variance increasing as
\beq
\alpha^{\rm CC} \sim \chi^{1/2}.
\eeq
As $\chi$ increases the closest-to-coherent (CC)
ensemble becomes more squeezed until eventually the linearization
leading to the above result breaks down. This indicates that is
not possible to physically realize an ensemble with a well-defined
coherent amplitude for a $\chi$ this large. This occurs when
$\alpha^{\rm CC} \sim \mu$, in other words $\chi \sim \mu^{2}$.
Note that this is larger than the critical value $\chi \sim
\mu^{3/2}$ at which the laser becomes incoherent, according to the
analysis of Sec.~\ref{seccoh}.

The situation is quite different in terms of the excess phase
diffusion parameter $\nu$.  As $\nu$  increases (with $\chi=0$)
the coherent state ensemble remains PR. This is true even when
$\nu > \mu^{2}$, the value at which the laser becomes incoherent,
as shown in Sec.~\ref{seccoh}. Moreover, phase diffusion tends to undo the
nonlinear effects of the self energy. In the limit $\nu \to \infty$,
the coherent state ensemble is PR for any finite
value of $\chi$.

\subsection{Interpretation}

In Ref.~\cite{Wis97}, the coherence condition for a laser, that the
output flux be much greater than the linewidth, was motivated by the
requirement that the laser have a well-defined phase. This follows
from the following argument. The laser phase remains fairly constant
over the coherence time (the reciprocal of the linewidth). However
this phase only has meaning if it can be measured, and this requires
a macroscopic field (i.e. many bosons) to be produced in the output
over one coherence time. As derived in Sec.~\ref{seccoh}, this
condition requires $\chi \ll \mu^{3/2}$ and $\nu \ll \mu^{2}$.

From the results of this paper there seems to be a problem with
this motivation for this definition of coherence. There are values
of $\chi$ between $\mu^{3/2}$ and $\mu^{2}$, and $\nu$ between
$\mu^{2}$ and $\infty$, for which the atom laser is not coherent
and yet for which it is possible to physically realize laser
states with well-defined coherent amplitudes.

The resolution of this problem is straight-forward for the case of
large $\nu$. The motivation in Ref.~\cite{Wis97} relied upon a
measurement of the phase {\em from the laser output}. By contrast, the
ensembles we have considered in this paper are physically realized by
monitoring {\em all} of the reservoirs of the laser. In particular,
that means monitoring the reservoirs that produce the excess phase
diffusion $\nu$. If we only allow for monitoring of the output of the
laser, the stochastic master equation will not preserve purity. After
linearization, the following equation results
\bqa
d{\rho} &=& (1/4) dt\left\{ {\cal D}[ x + iy]\rho +(1+\nu)  {\cal
D}[ x]\rho + {\cal D}[ y] \rho \right.
\nl{+}\left. {\cal H}[i (xy+yx)/2] +  {\cal
H}[-i\chi x^{2}] \right\}\rho \nn \\ && +\, (1/2) {\cal
H}[dW^{*}(t)  (x+iy)] \rho .
\label{lassme}
\eqa
Here there is only one stochastic term, from monitoring the laser
output. The best strategy for trying to realize states with
well-defined coherent amplitudes is clearly to measure the phase
quadrature of the output. This corresponds to $dWdW=-dt$.

Under these conditions, the differential equations for the
second-order moments of the conditioned state are
\bqa
\dot{\mu}_{20} &=& 2-2\mu_{20}-\mu_{11}^{2}, \label{int1}\\
\dot{\mu}_{11} &=& -\mu_{11}-\chi\mu_{20} -(\mu_{02}-1)\mu_{11}
\label{int2},\\
\dot{\mu}_{02} &=& -2\chi \mu_{11}+2+\nu -
(\mu_{02}-1)^{2}.\label{int3}
\eqa
If we set $\chi=0$, the steady-state solutions are
\bqa
\mu_{20} &=& 1, \\
\mu_{11} &=& 0 ,\\
\mu_{02} &=& 1+\sqrt{2+\nu}.
\eqa
In the limit of
large $\nu$ (which is the potential problem area), the phase
quadrature variance scales as $\nu^{1/2}$. The states lose their
coherent amplitude as the linearization breaks down at $\alpha =
\mu_{02} \sim \mu$. That is to say, at $\nu \sim \mu^{2}$. This is
precisely the regime identified in Sec.~\ref{seccoh} as that for which
the laser output loses its coherence.

Unfortunately (or perhaps fortunately from the point of view of
provoking new concepts), a similar analysis for large $\chi$ does not
hold. Instead, with $\nu = O(1)$ and $\chi \gg 1$ the solutions of
Eqs.~(\ref{int1})--(\ref{int3}) are
\bqa
\mu_{20} &\simeq& 2^{5/4}\chi^{-1/2}, \\
\mu_{11} &\simeq& -\sqrt{2} ,\\
\mu_{02} &\simeq& 2^{3/4}\chi^{1/2}.
\eqa
This is an extremely sheared state, with phase quadrature variance
scaling as $\chi^{1/2}$. It loses its well defined phase only for
$\chi \sim \mu^{2}$, which is the same scaling as found above when all
the reservoirs were unraveled. In particular, for $\mu^{3/2} < \chi <
\mu^{2}$, measuring the output has determined the phase of the laser
even though this should not be possible by the argument in
Ref.~\cite{Wis97} because the flux is less than the linewidth.

The difference between large $\nu$ and large $\chi$ can be
understood as follows. There are three Lindblad terms in the
linearized master equation (\ref{linlasme}). When $\nu=0$ they
are all of roughly the same size. Thus restricting the monitoring
to just one of the three reservoirs (the first one, the output)
has relatively little effect on the conditioned states. It is
much like monitoring all reservoirs, but with a reduced
efficiency. Indeed, the conditioned state in this case is not far
from a pure state, with $\mu_{20}\mu_{02}-\mu_{11}^{2}=2$
(compared to $1$ for a pure state). By contrast, with $\nu$ large
the phase diffusion Lindblad term is much larger than the other
two. Then if one is only able to monitor the output one is
necessarily losing most of the information about the system. This
leads to qualitatively different conditioned states, with much
reduced purity ($\mu_{20}\mu_{02}-\mu_{11}^{2} \simeq \sqrt{\nu}
\gg 1$).

The existence of the regime $\mu^{3/2} < \chi <
\mu^{2}$ where the laser output is incoherent,
but where the phase can in fact be determined
suggests that the concept of coherence time is more
subtle than the standard definition in terms of the first order
coherence function used in Ref.~\cite{Wis97} and in Sec.~\ref{seccoh}
above. The coherence time is also used to define whether or not the
laser beam is Bose degenerate, and, as discussed in
Ref.~\cite{Wis97}, the criterion is the same. That is, the output is
Bose degenerate if and only if many bosons come out ``with the same
phase'' (that is, within one
coherence time). Thus the present paradox has implications that go
beyond the present discussion, and impact on concepts like Bose
degeneracy as well, as will be discussed below.

\subsection{Conditional Coherence and Conditional Degeneracy}

One way to understand the above results is that the atom laser
for $\mu^{3/2} < \chi < \mu^{2}$ is ``conditionally coherent''.
The standard coherence condition $\chi < \mu^{3/2}$ can be
derived from the requirement that $\an{(\delta\phi)^{2}(t)} < 1$
at $t=1/\mu$, the time between atoms in the output. Here $(\delta
\phi)^{2}(t) \simeq y^{2}(t)/4\mu$ is the phase variance of the
state at time $t$, which was a coherent state at $t=0$. That this
implies the condition $\chi < \mu^{3/2}$ can be seen simply as
follows. For $\chi$ large and for a time as short as $1/\mu$, the
irreversible evolution can be ignored and the phase uncertainty
is due to the $Ca\dg a\dg a a$ Hamiltonian. For the linearized
theory, this turns into the Hamiltonian $\chi(x/2)^{2}$, where
$x$ is the amplitude quadrature. This causes the phase quadrature
to change as
\beq
y(t) = y(0) -\chi t x(0),
\eeq
where the mean frequency shift has been removed as has been
consistently done before. For a coherent state of zero mean phase we have
$\bar{y}(0)=0$, $\an{y(0)^{2}}=1$ and $\bar{x}(0)=0$, $\an{x(0)^{2}}=1$.
Thus for
$t=1/\mu$ we get
\beq
\an{[y(t)-\bar{y}(t)]^{2}}=1+\chi^{2}t^{2} = 1+\chi^{2}/\mu^{2}
\eeq
This is of order $4\mu$ (indicating the loss of coherence) for
$\chi \sim \mu^{3/2}$.

The coherent state is the most convenient state to use for this
calculation, as explained in Sec.~\ref{seccoh}. But of course it
is also possible to represent the atom laser as a mixture of
states with smaller amplitude uncertainty than a coherent state,
and, as we have seen, to physically realize such ensembles. The
average result must be the same, but the details are different.
Consider a minimum uncertainty pure state with
$V=\an{[x(0)-\bar{x}(0)]^{2}}=1/\an{y(0)^{2}}$, where the initial
mean phase has again been taken to be zero. The mean phase evolves
as
\beq
\bar{y}(t) = -\chi t \bar{x}(0),
\eeq
and the phase quadrature variance as
\beq
\an{[y(t)-\bar{y}(t)]^{2}} = 1/V + \chi^{2}t^{2}V
\eeq
To reproduce the stationary state which has a unit variance, we
must consider an ensemble of different values for $\bar{x}(0)$,
with mean zero and variance $1-V$. Thus the total phase variance
over the ensemble,
\bqa
 \an{[y(t)-\bar{y}(t)]^{2}} + {\rm E}\left[\bar{y}^{2}\right]
 &=& 1/V + \chi^{2}t^{2}V + \chi^{2}t^{2} {\rm E}[\bar{x}(0)^{2}] \nn \\
 &=& 1/V + \chi^{2}t^{2}
\eqa
cannot be less than that from a coherent state (with $V=1$).

In this picture, the increase in the phase uncertainty is the sum
of an intrinsic phase uncertainty increase and that due to an
uncertainty in the {\em frequency} of the field. The former is
due to an initial quantum uncertainty $V$ in the amplitude
quadrature, and the latter to a classical uncertainty $1-V$ in
the initial mean amplitude quadrature. The loss of coherence is
thus partly due to the addition of different interference terms
oscillating at different frequencies. For example, interfering
parts of the output field separated in time by $t$ would give a
different interference pattern depending on the frequency. Over a
time of order unity (the bare decay time), the mean amplitude
will sample all possible values so the frequency will also vary.
The average interference pattern measured over a time long
compared to this will thus be washed out due to the different
frequencies, and the experimenter would conclude that the output
was incoherent if $\chi^{2}t^{2} \sim \mu$ for $t\sim1/\mu$.

If, however, one {\em knows} (as the experimenter) the initial
mean amplitude $\bar{x}(0)$, then one knows what frequency to
expect in one's interference pattern. Then rather than simply
averaging the interference patterns over some long time, one
could correct for the mean frequency shift before doing the
average. Then the only contribution to the visibility of the
interference patter will be the intrinsic phase quadrature
variance
\beq
  \an{[y(t)-\bar{y}(t)]^{2}} = 1/V + \chi^{2}t^{2}V.
\eeq
From this {\em conditional} point of view, the laser output will
cease to be  coherent only when
\beq
  4\mu \sim 1/V + \chi^{2}(1/\mu)^{2}V.
\eeq
Solving for  $\chi$ gives
\beq
  \chi \sim \mu\sqrt{V^{-1}(4\mu-V^{-1})}.
\eeq
To maintain coherence for the largest possible $\chi$, we minimize
this with respect to $V$ to get
\beq
  \chi \sim 2\mu^{2}
\eeq
at $V \sim 1/2\mu$. This is the upper limit of the region $\mu^{3/2}
<\chi<\mu^{2}$ where a well-defined coherent amplitude is physically
realizable but the output is not coherent in the usual sense. Now we
can see that a physical realization giving the well-defined
coherent amplitude in this regime (such as that giving the
closest-to-coherent ensemble) is precisely what is required to
recover coherence, in a conditional sense.

The concept of conditionally coherent goes hand-in-hand with that
of conditionally Bose degenerate. Under the standard definition,
the atom laser output in the regime $\mu^{3/2}<\chi<\mu^{2}$ is
not Bose degenerate. Specifically, there is no mode that can be
identified {\em a priori} in the output and that has a large mean
occupation number. But under an unraveling of the atom laser
dynamics, such a mode can be identified in this regime: it is a
mode corresponding to the frequency which can be inferred from
the knowledge of the amplitude of the condensate. As with the
case of conditional coherence, a new mode will have to be chosen
after a short time, since the frequency explores the full range
on a time scale of order unity. But at a particular instant of
time, the knowledge obtained from monitoring the reservoirs of
the system (or even just the output, as seen above) is sufficient
to allow a highly-occupied mode to be identified.

We can perhaps clarify the concept of conditional Bose degeneracy as
follows. Consider a system with $N$ modes, and $N$ particles. The
multiparticle state
\beq
\rho=\ket{N_{1},0_{2},0_{3},\ldots,0_{N}}
\bra{N_{1},0_{2},0_{3},\ldots,0_{N}}
\eeq
is clearly  Bose-degenerate, just as the state
\beq
\rho = \ket{1_{1},1_{2},1_{3},\ldots,1_{N}}
\bra{1_{1},1_{2},1_{3},\ldots,1_{N}}
\eeq
is not. But what about the state
\bqa
\rho &=& N^{-1} \sum_{m=1}^{N} \ket{\ldots
0_{m-2},0_{m-1},N_{m},0_{m+1},0_{m+2}\ldots} \nl{\times} \bra{\ldots
0_{m-2},0_{m-1},N_{m},0_{m+1},0_{m+2}\ldots}\,?
\eqa
The mean occupation number of any mode is clearly one, so it is not
Bose degenerate in the usual sense. But also clearly if one had access
to this state then after finding a single particle, one would know in
what state the remaining $N-1$ particles would lie. Thus the state
would have become {\em conditionally Bose degenerate}. We believe
that the above state is a good toy description of a short section of
the  output of an
atom laser in the interesting regime of $\mu^{3/2}<\chi<\mu^{2}$,
where the different modes represent different frequencies.

Finally, it is interesting to note that by employing feedback
based on QND atom number measurements, it is possible (within the
current atom laser model) greatly to reduce the linewidth
\cite{WisTho01}. Specifically, the linewidth may be reduced by a
factor of order $\mu^{1/2}$, and the coherence (in the
conventional sense) of the laser extended from $\chi \alt
\mu^{3/2}$ to $\chi \alt \mu^{2}$. This is not quite an exact
parallel with the above results, because the feedback is based on
a measurement that adds extra phase diffusion ($\nu$) term, that
is not required in the above analysis. (This QND measurement is
introduced because it is a number measurement, and so is more
easily realized than the phase-sensitive measurement necessary in
the above analysis.) Nevertheless, it still illustrates the
general principle stated in Ref.~\cite{Wis93}, that ``the
practical significance of [conditional analyses] is that
conditioning is realized by feedback.''

\subsection{Experimental Implications} \label{sec:expt}

It is clear that many interesting questions relating to the
coherence of an atom laser, the physical realizability  of a
coherent state ensemble, the coherence of the output, and the
conditional coherence of the output, depend upon the value of
$\chi$. This prompts the question:   what value has this
parameter in experimental atom lasers?  As discussed in the
introduction, a number of experimental groups have realized
Bose-Einstein condensates with output coupling
\cite{Mew97,And98,Hag99,Blo99}. A CW atom laser would have to
incorporate a mechanism for replenishing the condensate so that
the output coupling could continue indefinitely. Nevertheless we
can take these experiments as a possible indication for the
parameter regime in which an atom laser may work.  The figures
below are derived by setting the bare linewidth $\kappa$ of the
laser equal to the reciprocal of the lifetime of the condensates
in the experiment, and the mean atom number $\mu$ equal to the
initial occupation number of the condensate. The excess phase
diffusion $\nu$ we have ignored, and we have calculated $\chi$
using \erf{selfenergy} and \erf{nuchi}.

Most current experiments work in the regime where the ratio of
the kinetic energy to the interaction energy is very small
\cite{BayPet96}:
\beq
\left( \frac{\hbar}{64 \pi^2  m\omega \mu^2 a_s^2} \right)^{2/5}
\ll 1.
\eeq
Here $m$ is the atomic mass, $\omega$ is the mean trap frequency,
and $a_s$ is the scattering length as in \erf{selfenergy}. In this
regime the Thomas-Fermi approximation can be made, allowing us to
evaluate $\chi$ analytically as
\beq
\chi = \frac{4}{7\kappa}\,\left( \frac {225 \mu^2 m \omega^6
a_s^2} {\hbar} \right)^{1/5}.
\eeq
The values of $\chi$ using the parameters of three recent
experiments are compared in Table I.  The MIT experiment
\cite{Mew97} represents the first ``pulsed atom laser'', a
quasicontinuous output coupling \cite{Hag99} was demonstrated at
NIST, and the MPQ experiment \cite{Blo99} demonstrated a continuous
output coupling.

\vspace{4mm}
\begin{center}
  \centering
   \begin{tabular}{l|r|r|r|r} \hline
                              & MIT    &   MPQ   &   NIST  &  Proposed \\
    \hline
    $\chi$                    & 910    &  1800   &  50     &    990   \\
    $IT$                      & $4.1\times 10^6$ & $2.1\times 10^5$
                              & $5.7\times 10^6$ & $2.9\times 10^5$   \\
    $I/\ell$                  & $6.0\times 10^7$ & $4.0\times 10^5$
                              & $8.0\times 10^8$ &  $2.0\times 10^6$  \\
    $\kappa/\ell$             &  12    &  0.57   &  810    &     2.0  \\
    $\omega_{\rm min}/\kappa$ &  1.1   &  4.8    &  0.8    &     22   \\
    $\omega_{\rm min}/\ell$   &  14    &  2.7    &  640    &     44   \\
    \hline
   \end{tabular}

\small
Table I. Parameters for recent (and proposed) atom laser
experiments at various institutions
\end{center}

All $\chi$ values are in the $\chi \gg 1$ regime on which we have
concentrated in this paper. Thus if these experiments could be run
with the same output coupling but with continuous replenishment of
the condensate, the closest-to-coherent ensemble that could be
physically realized would be highly amplitude squeezed. From
\erf{gammaCC}, with $\chi=1000$ the standard deviation of the
amplitude-quadrature of these states would be about $0.2$,
compared to $1$ for coherent states. Thus it seems that it is
wrong to think of an atom laser as being in a coherent state.

Despite the banishing of the coherent state description, truly
continuous versions of the  experiments analyzed above would
produce an unconditionally coherent (Bose degenerate) output. That
is because the calculated values of $\chi$ are always much less
than $\mu^{3/2}$, so that \erf{cohcon1} above is satisfied.
Interestingly, we can recast this condition in terms of the output
flux $I=\kappa\mu$ (atoms per unit time) as
\beq
    I \gg 1.61 \, \omega \left(\frac {a_s^4 \omega m^2
    \kappa}{\hbar^2}\right ) ^{1/11}
\eeq
This inequality depends very weakly on the dimensionless quantity
in brackets because of the $11$th root. For the above three
experiments this $11$th root averages to $0.16$, and ranges only
from $0.13$ to $0.21$. Hence we can state the coherence condition
for an atom laser in terms essentially independent of the species
and decay time as $I \gg 0.26 \omega$ or
\beq
    I \gg T^{-1} .
\eeq
That is, there should be many atoms emitted into the laser beam
per oscillation period $T=2\pi/\omega$ of the trap. This is such a
simple rule of thumb that it should be useful, but it must be
remembered that there is no direct physical connection between the
flux and the trap frequency. This result is simply a numerical
coincidence arising from the various physical parameters for
atomic Bose-Einstein condensation in typical traps.  The second
row of the table shows that this condition is clearly satisfied
for the parameters of the three experiments and this suggest that
the output field of our model atom laser would be degenerate.

The actual degree of degeneracy $D$ of the output field, that is
the number of atoms per output frequency mode, is given by the
quotient $I/\ell$ of the flux $I$ and the linewidth $\ell$. The
linewidth for the atom laser model we are considering is given in
Ref.\ \cite{WisTho01} as
\bqa
  \ell \simeq \left\{
      \begin{array}{ll}
         \kappa(1+\chi^2)/2\mu & \ {\rm for }\ \chi < \sqrt{8\mu/\pi}\\
         2\kappa\chi/\sqrt{2\pi\mu} & \ {\rm for }\ \chi > \sqrt{8\mu/\pi}
      \end{array}\right.
      \label{linewidth}
\eqa
The third row of the table shows that, for the same parameters as
the experiments, the output field of the atom laser model is
highly degenerate.

It is interesting to compare the linewidth of the output field
$\ell$ with the bare cavity linewidth $\kappa$.  The action of the
pump tends to reduce the linewidth far below $\kappa$ in the same
manner of an optical laser [the $\chi\to 0$ limit of
\erf{linewidth}]. In an atom laser, however, the nonlinearity
converts intensity fluctuations into phase fluctuations and this
tends to broaden the linewidth [the $\chi\to\infty$ limit of
\erf{linewidth}].  Table I shows a range of values of the ratio
$\kappa/\ell$ from below unity (line broadening) to well above
unity (line narrowing) for the parameters of the experiments. We
can write $\kappa=I/\mu$ and $\ell=I/D$ and so the ratio
$\kappa/\ell=D/\mu$ is also the ratio of the number of atoms per
output frequency mode to the steady state population in the
cavity. Significant line narrowing therefore leads to $D\gg\mu$,
that is, many more atoms per output mode than in the condensate.

Our analysis assumes that we can treat the atomic condensate as a
single atomic field mode.  We now show how this assumption can be
justified with realistic experimental conditions. Only a single
mode is needed if the condensate is, at most, only weakly coupled
to the quasiparticle modes. There are two important ways in which
this coupling can arise.  One is due to the fact that the spatial
form of the quasiparticle modes depends on the number of atoms in
the condensate and so fluctuation in the condensate number will
cause an overlap between condensate and quasiparticle modes.
However, provided the fluctuations in the condensate atom number
occur on a time scale much longer that the dynamics of the
condensate and quasiparticle modes, the system will evolve
adiabatically and remain in the condensate mode.  Thus the first
requirement for minimal coupling to the quasiparticle modes is
\beq
  \omega_{\rm min}/\kappa \gg 1
  \label{singlemode1}
\eeq
where $\omega_{\rm min}$ is the lowest of the trap frequencies.
The other coupling mechanism is due to the linewidth of the
condensate mode.  In order to avoid adiabatic exchange of atoms
between condensate mode and quasiparticle modes, we need the
linewidth to be much smaller than the spacing between the
condensate mode and first excited mode.  This difference is simply
the lowest trap frequency $\omega_{\rm min}$ \cite{Str96}. Hence the second
requirement for a single mode analysis is
\beq
   \omega_{\rm min}/\ell  \gg 1\ .
  \label{singlemode2}
\eeq We have tabulated figures for these parameters in Table I for
the three experiments and included further data for a proposed
experiment. The three experiments are clearly not operating in the
single mode regime as $\omega_{\rm min}/\kappa$ or $\omega_{\rm
min}/\ell$ or are order unity.  So besides not being continuously
pumped, the experiments also do not satisfy the single mode
criteria of our model and thus require a pulsed, multimode
analysis such as that of Ref.\ \cite{RGH01}.  However it would not
be difficult to achieve single mode operation by selecting
different, but experimentally reasonable, parameters. For example,
the last column in the table shows the values for a Sodium atom
laser in a symmetric trap with frequency $\omega=\omega_{\rm
min}=2\pi\times 25 {\rm Hz}$, output coupling rate $\kappa=7{\rm
s}^{-1}$ and mean atom number of $\mu=10^6$. Both conditions
\erf{singlemode1} and \erf{singlemode2} are satisfied and so the
coupling would be minimal in this case.

\subsection{Closing Remarks}

It is fitting to end by referring to the very beginning, that is,
the title of our paper. What does the physical realizability of
ensembles of pure states say about atom lasers, coherent states
and coherence?

First, they establish a basis on which it is
possible to objectively discuss the existence of coherent states
as the state for an atom laser.

Second, they show that these coherent states can only exist (that
is, be physically realized) for $\chi=0$ (that is, in the total
absence of interactions between the atoms).

Third, the existence of pure states {\em close} to coherent states
requires $\chi \ll 1$, which is a much stronger condition than the
$\chi \ll \mu^{3/2}$ needed for the laser output to be coherent
(Bose degenerate).

Fourth, the existence of states with well-defined coherent
amplitude (that is, with phase variance small compared to unity)
requires $\chi \ll \mu^{2}$, a far weaker condition that that
needed for realizing coherent states, and also weaker than that
required for output coherence.

Fifth, in the regime $\mu^{3/2} \alt \chi \ll \mu^{2}$, a new concept
of coherence (and Bose degeneracy) pertains, that of {\em
conditional coherence} (or {\em conditional Bose degeneracy}). In
this regime, knowing which member of physically realizable
ensemble one has at a given point in time allows the coherence to
be demonstrated, where it could not be in the absence of that
knowledge.

Sixth, unlike $\chi$, excess phase diffusion $\nu$ does not destroy the
physical realizability of the coherent state ensemble (for $\chi=0$),
and in fact
makes it easier to approach this ensemble for finite $\chi$.

Seventh, the existence of a regime ($\nu \agt \mu^{2}$) in which the
laser output is incoherent but an ensemble of states with well
defined coherent amplitudes (indeed, coherent states) is physical
realizable, does not require a new concept of coherence. Rather,
by restricting the measurement of the atom laser to the monitoring
of its output beam itself, the physical realizability of such an
ensemble is restricted to the coherent-output regime $\nu \ll
\mu^{2}$.

\acknowledgments

We thank Dr. J. Ruostekoski for discussions regarding the validity
of the single mode analysis. H.M.W. is supported by the Australian
Research Council.

\end{multicols}

\rule[\baselineskip]{\textwidth}{0.4pt}

\begin{multicols}{2}
\begin{figure}
\psfig{figure=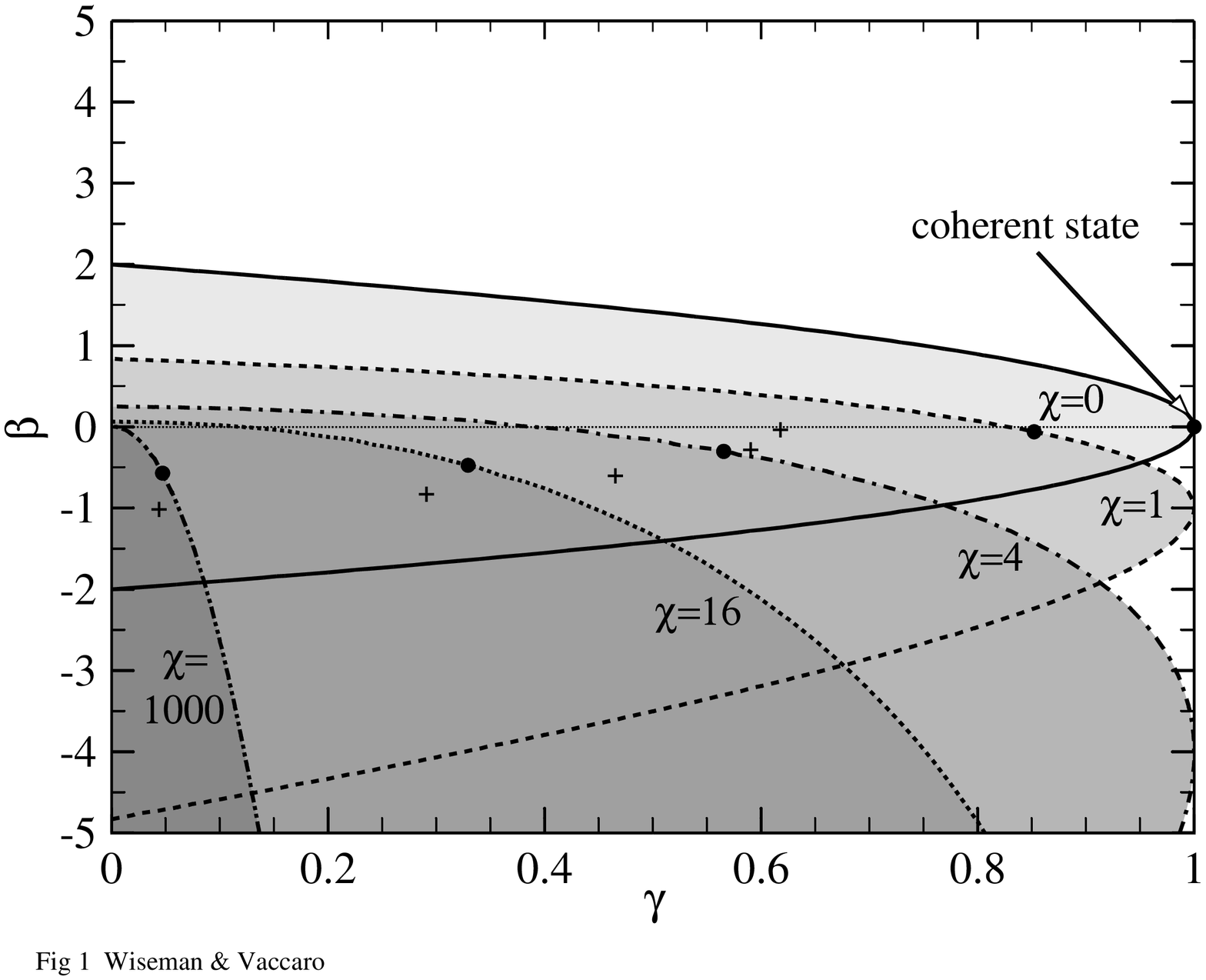,width=80mm,angle=-90,bbllx=17mm,bblly=30mm,bburx=193mm,bbury=263mm,clip=}
\caption{\narrowtext Representation of physically-realizable
ensembles, arising from general continuous Markovian unravelings
(CMU), for $\nu=0$ and various values of $\chi$. The shaded
regions represent values of $\gamma$ and $\beta$ (and thus
$\alpha=(1+\beta^2)/\gamma$) that can be realized by monitoring.
The progressively-darker shaded regions correspond to values of
$\chi$ of 0, 1, 4, 16 and 1000 and are bounded by solid, dashed,
dash-doted, dotted and dash-dot-doted curves, respectively. The
$\gamma$, $\beta$ value of the closest-to-coherent (CC) ensemble
in each region is marked as a filled circle on the boundary. The
crosses mark the $\gamma$, $ \beta$ values of the quantum-state
diffusion (QSD) ensembles for the same set of $\chi$ and $\nu$
values, with the $\chi$ values reducing from left to right.
    \protect\label{fig1}}
\end{figure}

\begin{figure}
\psfig{figure=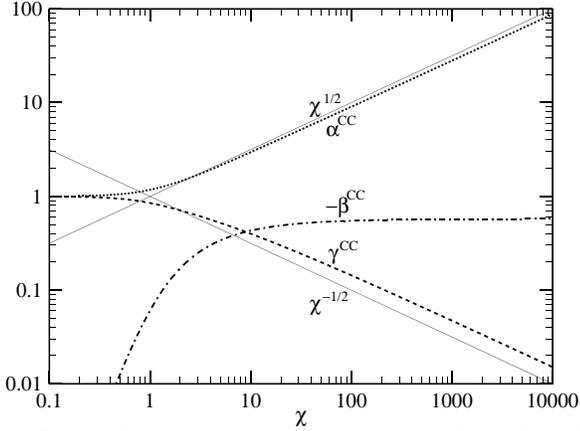,width=80mm,angle=-90,bbllx=17mm,bblly=30mm,bburx=193mm,bbury=273mm,clip=}
\caption{\narrowtext The parameters for the physically-realizable
ensemble that is closest to a coherent ensemble (CC) as a function
of $\chi$ with $\nu=0$. The ensembles arise from general
continuous Markovian unravelings. These parameters are the phase
quadrature variance $\alpha^{\rm CC}$ (dotted line), the
amplitude-quadrature variance $\gamma^{\rm CC}$ (dashed line) and
the covariance $\beta^{\rm CC}$ (dash-dot line) for the members of
this ensemble.  Also shown for comparison are thin solid curves
representing $\chi^{1/2}$ and $\chi^{-1/2}$.
    \protect\label{fig2}}
\end{figure}

\begin{figure}
\psfig{figure=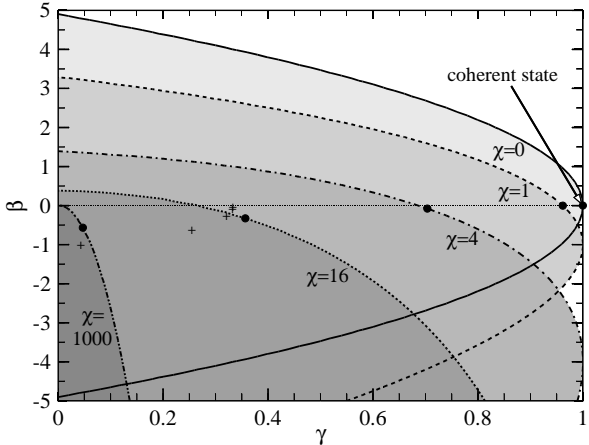,width=80mm,angle=-90,bbllx=17mm,bblly=30mm,bburx=193mm,bbury=263mm,clip=}
\caption{\narrowtext Representation of physically-realizable
ensembles similar to Fig.~1 but for $\nu=10$.
    \protect\label{fig3}}
\end{figure}

\begin{figure}
\psfig{figure=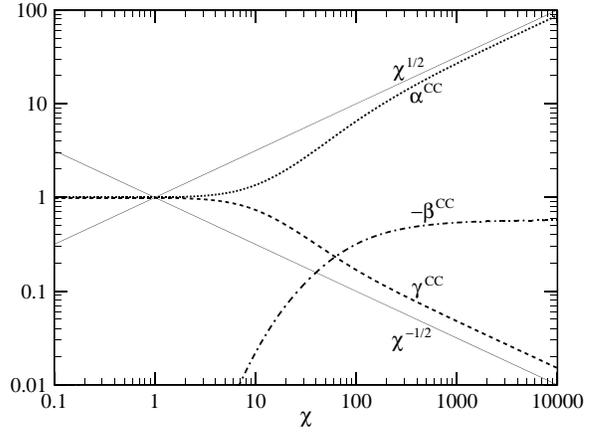,width=80mm,angle=-90,bbllx=17mm,bblly=30mm,bburx=193mm,bbury=273mm,clip=}
\caption{\narrowtext The parameters for the closest-to-coherent
physically realizable (CC PR) ensemble as a function of $\chi$
similar to Fig.~2 but here with $\nu=100$. The excess phase
diffusion allows the realization of states very close to coherent
states until $\chi\sim\nu^{1/2}$
    \protect\label{fig4}}
\end{figure}

\begin{figure}
\psfig{figure=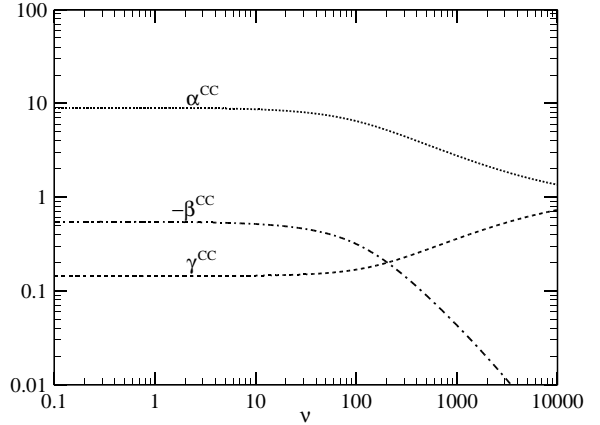,width=80mm,angle=-90,bbllx=17mm,bblly=30mm,bburx=193mm,bbury=273mm,clip=}
\caption{\narrowtext The parameters of the closest-to-coherent
physically-realizable (CC PR) ensemble similar to Fig.~2 but here
as a function of $\nu$ and with $\chi=100$.
      \protect\label{fig5}}
\end{figure}

\begin{figure}
\psfig{figure=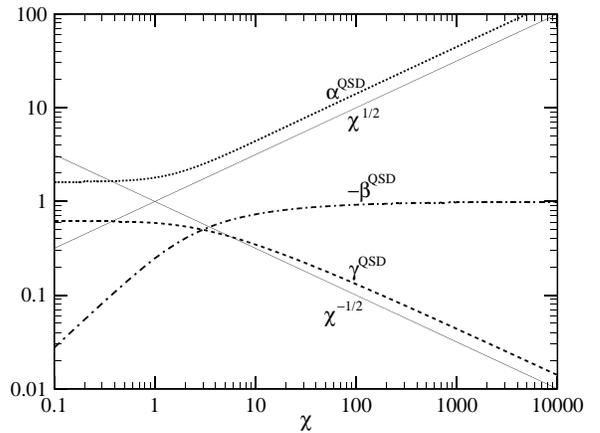,width=80mm,angle=-90,bbllx=17mm,bblly=30mm,bburx=193mm,bbury=273mm,clip=}
\caption{\narrowtext The parameters of the ensemble arising from
quantum state diffusion (QSD) as a function of $\chi$ with
$\nu=0$.  The labeling follows Fig.~2.
    \protect\label{fig6}}
\end{figure}

\begin{figure}
\psfig{figure=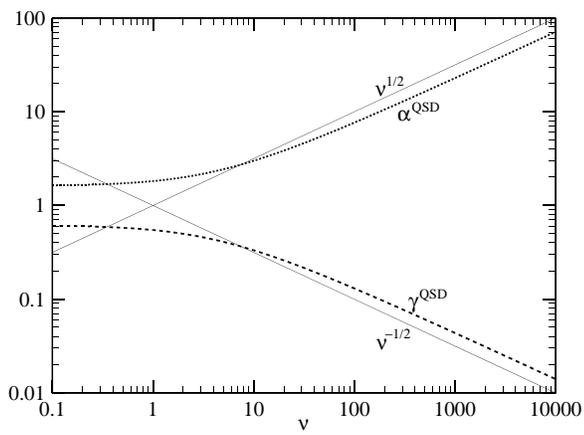,width=80mm,angle=-90,bbllx=17mm,bblly=30mm,bburx=193mm,bbury=273mm,clip=}
\caption{\narrowtext Similar to Fig.~6 but as a function of $\nu$
with $\chi=0$.  The thin solid curves represent values of
$\nu^{1/2}$ and $\nu^{-1/2}$.
    \protect\label{fig7}}
\end{figure}
\end{multicols}

\end{document}